\documentclass[preprint]{aastex}

\shorttitle{{\it AKARI} Observation of NEP supercluster at z=0.087}

\shortauthors{Jongwan Ko et al.}

\begin{document}

\title{{\it AKARI} Observation of the North Ecliptic Pole (NEP) Supercluster at z = 0.087:
       mid-infrared view of transition galaxies}

\author{Jongwan Ko\altaffilmark{1,2,3,4} , Myungshin Im\altaffilmark{3,4}, 
 Hyung Mok Lee\altaffilmark{3}, Myung Gyoon Lee\altaffilmark{3}, Seong Jin Kim\altaffilmark{3}, 
 Hyunjin Shim\altaffilmark{5}, Yiseul Jeon\altaffilmark{3,4}, Ho Seong Hwang\altaffilmark{6}, 
 Christopher N. A. Willmer\altaffilmark{7}, Matthew A. Malkan\altaffilmark{8},
 Casey Papovich\altaffilmark{9}, Benjamin J. Weiner\altaffilmark{7}, Hideo Matsuhara\altaffilmark{10}, 
 Shinki Oyabu\altaffilmark{11}, Toshinobu Takagi\altaffilmark{10}}

\altaffiltext{1}{Yonsei University Observatory, Yonsei University, Seoul 120-749, Korea}
\altaffiltext{2}{Korea Astronomy and Space Science Institute, Daejeon 305-348, Korea}
\altaffiltext{3}{Astronomy Program, Department of Physics \& Astronomy, FPRD, Seoul National University, Seoul 151-742, Korea}
\altaffiltext{4}{Center of the Exploration of the Origin of the Universe (CEOU), Seoul National University, Seoul, Korea}
\altaffiltext{5}{Spitzer Science Center, California Institute of Technology, MS 220-6, Pasadena, CA 91125}
\altaffiltext{6}{Service d'Astrophysique, CEA Saclay, F-91191 Gif-sur-Yvette, France}
\altaffiltext{7}{Steward Observatory, University of Arizona, 933 N. Cherry Avenue, Tucson, AZ 85721, USA}
\altaffiltext{8}{Department of Physics and Astronomy, University of California at Los Angeles, CA 90095, USA}
\altaffiltext{9}{George P. and Cynthia W. Mitchell Institute for Fundamental Physics and Astronomy, Department of Physics, Texas A\&M University, College Station, TX 77843, USA}
\altaffiltext{10}{Institute of Space and Astronautical Science, Japan Aerospace Exploration Agency, Kanagawa 229-8510, Japan}
\altaffiltext{11}{Graduate School of Science, Nagoya University, Furo-cho, Chikusa-ku, Nagoya, Aichi 464-8602, Japan}

\email{jwko@yonsei.ac.kr}

\begin{abstract}

We present the mid-infrared (MIR) properties of galaxies within a supercluster in the North 
Ecliptic Pole region at z$\sim$0.087 observed with the AKARI satellite. We use data from the AKARI 
NEP-Wide (5.4 deg$^{2}$) IR survey and the CLusters of galaxies EVoLution studies 
(CLEVL) mission program. We show that near-IR (3 $\mu$m)--mid-IR (11 $\mu$m) color 
can be used as an indicator of the specific star formation rate and the presence of 
intermediate age stellar populations. From the MIR observations, we find that red-sequence 
galaxies consist not only of passively evolving red early-type galaxies, but also of 
1) ``weak-SFG" (disk-dominated star-forming galaxies which have star formation rates lower 
by $\sim$ 4 $\times$ than blue-cloud galaxies), and 
2) ``intermediate-MXG" (bulge-dominated galaxies showing stronger MIR dust emission 
than normal red early-type galaxies). Those two populations can be
a set of transition galaxies from blue, star-forming, late-type galaxies evolving into red, 
quiescent, early-type ones. We find that the weak-SFG are predominant at intermediate 
masses ($10^{10}M_{\odot} < M_{*} < 10^{10.5}M_{\odot}$) and are typically found in local 
densities similar to the outskirts of galaxy clusters. As much as 40\% of the supercluster 
member galaxies in this mass range can be classified as weak-SFGs, but their proportion
decreases to $<$ 10\% at larger masses ($M_{*}$ $>$ 10$^{10.5}$ $M_{\odot}$) at any galaxy density. 
The fraction of the intermediate-MXG among red-sequence galaxies at 
$10^{10}M_{\odot} < M_{*} < 10^{11}M_{\odot}$ also decreases as the density and mass increase. 
In particular, $\sim$42\% of the red-sequence galaxies with early-type morphologies are classified 
as intermediate-MXG at intermediate densities. These results suggest that the star formation activity 
is strongly dependent on the stellar mass, but that the morphological transformation is mainly 
controlled by the environment.

\end{abstract}


\section{INTRODUCTION}

One of the leading factors that can strongly influence galaxy evolution is the environment.
Observationally, it has been known that the environment plays an important role in shaping 
galaxy properties (see Blanton \& Moustakas 2009 for a review). The
morphology-density relation (MDR) was first described by Dressler
(1980), who found a strong correlation between the morphological type
fraction and local galaxy surface density, where, for increasing local density,
the fraction of elliptical galaxies increases, while the spiral fraction decreases.
Since then, a number of studies have reported that galaxy properties such as colors, 
and star formation activity (SFA) are also strongly dependent on the local density 
(e.g., Park \& Hwang 2009). MDR was also found at z$\sim$1 (e.g., Postman et al. 2005; 
Hwang \& Park 2009). Similarly, the fraction of blue, star-forming (SF) galaxies decreases 
as the density increases, which is known as the color-density relation 
(CDR, e.g., Lewis et al. 2002; Balogh et al. 2004; Koopmann \& Kenney 2004). 
More recently, using the galaxies in the Galaxy Zoo project (Lintott et al. 2008), 
Bamford et al. (2009) studied the dependence of galaxy color and morphology on environment, 
and showed that galaxy color has a much stronger dependency on environment than morphology 
at fixed stellar mass. This is consistent with the result of Blanton et al. (2005), which 
suggests that the color seems to be very sensitive to the local density.

Besides environment, there is also a clear tendency for massive galaxies at low redshift 
to be red, quiescent, and have early-type morphologies, indicating that
the galaxy properties also correlate with stellar mass. Kauffmann et al. (2003), 
using the Sloan Digital Sky Survey (SDSS; York et al. 2000) data, showed that 
low-redshift galaxies divide into two distinct populations at a stellar mass of 
3 $\times$ 10$^{10}$ M$_{\odot}$: lower-mass galaxies have young stellar populations 
and low concentration indices typical of disk systems.
Baldry et al. (2006) also found that the color-mass relations do not depend strongly 
on environment, while the fraction of red galaxies depends both on mass and environment. 
They also found that models with internally driven feedback mechanisms can explain the 
observed properties better. 
In a different study, Bundy et al. (2006) investigated the mass-dependent evolution 
of galaxies for 0.4 $<$ z $<$ 1.4, and found that there is no significant correlation 
between environment and the ``downsizing'' trend -- in which massive galaxies
are older because their SFA finished earlier (Cowie et al. 1996). This suggests 
that quenching of star formation in massive galaxies is primarily internally driven. 
This idea is also consistent with recent findings using z $\sim$ 1 samples, that the 
SFA is most strongly driven by stellar mass (e.g., Peng et al. 2010; Li et al. 2011; 
Sobral et al. 2011).

Therefore, the question of whether mass or environment is a main driving factor for changes
in color and morphology, and how they have affected the evolution of galaxies remain one of 
the controversial issues of galaxy evolution. We cannot track individual galaxies as they evolve with 
time from their birth. 
In the local universe we can only see evolutionary snapshot pictures of galaxies, from which 
we infer their evolution indirectly. Galaxies in the transition phase from blue 
to red and from spiral to spheroidal are particularly useful targets to study the 
evolutionary mechanisms. In this paper, we focus on these transition populations, 
and explore their dependence on mass and environment.

Clusters and Superclusters are excellent astrophysical laboratories of galaxy evolution. 
These structures provide a wide range of environments, ranging from low density at the 
outskirts to the high density regions at the cluster centers. 
They also contain a wide range of galaxy masses -- from 
low mass dwarf galaxies ($M_{*}$ $\sim$ 10$^{7}$ M$_{\odot}$) to 
the most massive galaxies ($M_{*}$ $\sim$ 10$^{13}$ M$_{\odot}$).
In addition to the MDR and the CDR mentioned earlier, the number of SF galaxies 
(blue galaxies) in clusters increases toward higher redshift (Butcher-Oemler effect; 
Butcher \& Oemler 1984), and the number of S0 galaxies seems to decline rapidly with 
redshift, with a corresponding increase in the blue, spiral galaxies (e.g., Dressler et al. 1997;
Poggianti et al. 2001). At higher redshift, the star formation rates (SFRs) in high-density 
environments seem to be steadily increasing (Elbaz et al. 2007). The lack of SF galaxies in 
cluster centers, and the possible transformation of blue galaxies into red, S0 galaxies prompted 
the suggestion of  many physical mechanisms to carry out this transformation, whose cause is 
either gravitational (tidal interaction), or hydrodynamic (e.g., ram-pressure stripping of 
interstellar gas; removal of halo reservoir halting the gas supply and quenching star formation; 
see Boselli \& Gavazzi 2006 and Park \& Hwang 2009).

Many researchers have used optical colors or spectral features as a proxy of SFA, 
to study the galaxy properties in cluster environments. Stellar masses of galaxies
are usually obtained by fitting the spectral energy distributions (SED)
measured by UV and optical photometry. Large datasets of galaxies,
such as the SDSS, have been used to understand the galaxy evolution in
cluster environments (e.g., Blanton \& Moustakas 2009; Park \& Hwang 2009). However, an obvious 
disadvantage of this approach using the UV/optical light is dust obscuration. 
The interstellar medium in galaxies has dust, which absorbs the UV light from 
stars (mostly coming from young, hot stars). This leads to significant extinction 
of the UV and optical light, complicating the interpretation of the SFA. Fortunately, 
the absorbed UV/optical light is re-emitted in the infrared, where one can obtain an 
unobscured view of the SFA. Earlier studies of the SFA in clusters were made with ISO 
(e.g., Boselli et al. 1998; Biviano et al. 2004). More recently, Spitzer has revealed 
the SFA in clusters in the IR, finding a few clusters with exceptionally high SFRs, and 
the rapid evolution of the SFA as a function of redshift (Bai et al. 2009). Nevertheless, the 
study of the SFA in clusters, as a function of environment, has been limited.
 
In cluster environments, early-type galaxies follow a tight color-magnitude relation 
(red-sequence), indicating their stellar population is homogeneously old and passively 
evolving (e.g., Bower et al. 1992; Kodama \& Arimoto 1997). However, early-type galaxies 
do not contain homogeneous stellar populations when we examine them at different wavelengths,
particularly in the IR. Previous IR observations showed that there are some early-type galaxies 
with excess far-IR (FIR) emission (Knapp et al. 1989), and mid-IR (MIR) emission (Knapp et al. 1992; 
Xilouris et al. 2004). Recently, Clemens et al. (2009), using Spitzer-IRS peakup images 
(16 $\mu$m), found that about 32\% of the early-type galaxies in the Coma cluster have excess 
flux over photospheric emission in the MIR. Bressan et al. (2006) also detected MIR emission in 
early-type galaxies with Spitzer, showing a wide emission feature around 10 $\mu$m and another 
broad feature near 18 $\mu$m. Unusual polycyclic aromatic hydrocarbons (PAHs) are also detected 
in the NIR/MIR spectra of nearby early-type galaxies (Kaneda et al. 2005, 2008; Lee et al. 2010; 
Vega et al. 2010; Panuzzo et al. 2011). This suggests that they are associated with 
intermediate-age stellar populations, formed in a post-starburst phase.

The IR emission from early-type galaxies is attributed either to the Rayleigh-Jeans tail of the 
stellar photosphere, or to circumstellar dust around evolved stars in the Asymptotic Giant Branch (AGB). 
Theoretical works show that the MIR-excess emission of AGB stars is well-correlated with stellar age 
(Piovan et al. 2003), and several studies suggest that MIR-excess is a good age indicator (Temi et al. 
2005; Ko et al. 2009; Shim et al. 2011). Indeed, the MIR-excess emission can be useful in tracing the 
past SFA, because other mean stellar age indicators cannot discriminate the emission 
of these stars.

In summary, earlier studies find that IR data are useful and perhaps critical in some cases to 
(i) obtain an unobscured view of the SFA and (ii) trace recent SFA in early-type or red galaxies. 
However, the previous studies of IR properties of cluster galaxies have been mostly limited to 
the derivation of global properties of clusters, or examining a limited number of individual galaxies. 
In addition, there are few studies focusing on the galaxy environment on much larger scales (such as 
across a supercluster). Therefore, it is necessary to study the effects of environment and/or mass 
on the IR properties of galaxies in clusters and superclusters.

Batuski \& Burns (1985) first discovered the large-scale structure in the North Ecliptic Pole (NEP) 
region as an association of six clusters of galaxies, using a percolation analysis of clusters in 
the Abell's (1958) catalog. Subsequently,  Burg et al. (1992), using early observations with the 
ROSAT satellite, reported five X-ray clusters and groups at 0.08$<$z$<$0.09 within 1.5$^{\circ}$ 
of the NEP. With the deepest exposure of the ROSAT All-Sky Survey, Mullis et al. (2001) found an 
extended large-scale structure in the NEP region at 0.07$<$z$<$0.1 -- the NEP supercluster
-- which consists of eight Abell clusters (A2255, A2295, A2301, A2304, A2308, A2311, A2312, and A2315). 
These X-ray detected clusters have luminosities in the range of 
(0.2$-$3.6) $\times$ $10^{44}$ $h^{-2}_{50}$ ergs $s^{-1}$ (0.5 $-$ 2.0 keV). In addition, they 
found three new clusters in this X-ray luminosity range, which are not contained in the optical 
cluster catalog, and ten groups of galaxies with X-ray luminosities in the range of (2$-$9) 
$\times$ $10^{42}$ $h^{-2}_{50}$ ergs $s^{-1}$ (0.5 $-$ 2.0 keV).

Thanks to the AKARI IR Space Telescope (Murakami et al. 2007) and its Sun-synchronous orbit, 
we carried out an IR survey in the NEP region (5.4 deg$^{2}$) as part of the AKARI NEP-Wide survey
(ANWS, see Figs.1 and 3). The main advantage of AKARI against previous IR satellites is the continuous 
wavelength coverage from the near- to mid-IR (2 to 24$\mu$m), especially at 11 and 15 $\mu$m, which 
allows detecting the MIR-excess dust emission from circumstellar matter around AGB stars. Therefore, 
AKARI observations are well-suited for the study of MIR-excess from AGB stars, as well as obscured SFA. 
The ANWS covers intermediate- and low-density regions at 0.07$<$z$<$0.1 where the NEP supercluster is 
located. To compare the galaxy properties in these regions with those in high-density regions, we include 
the galaxy data of A2255 that was also observed by AKARI as part of the mission program of CLusters 
of galaxies EVoLution studies (CLEVL; Im et al. 2008; Lee et al. 2009b). The AKARI observations for 
the A2255 field were carried out using 6 IRC filters (Shim et al. 2011). The depth of the A2255 data 
is comparable to that of the ANWS. Thus, the uniformity of our AKARI IR data across a range of environments 
(a rich cluster, three groups, and lower density regions) allows us to 
directly investigate the galaxy populations and their environmental dependence.

The primary goal of this study is to understand how environment and mass affect the evolution of 
galaxies, focusing on the quenching of the SFA, and morphological transformation.
To do this, we focus on transition populations (red-sequence galaxies with various MIR properties), 
taking advantage of the multi-wavelength (UV-to-MIR) data for galaxies in the NEP supercluster at 
0.07 $<$ z $<$ 0.1. Most importantly, the 11$\mu$m flux traces not only the mean stellar age and 
the specific SFR (SSFR) of SF galaxies, but also the presence of intermediate age stellar populations, 
detecting even tiny amounts of past star formation in early-type galaxies. 

Throughout this paper, we use \textit{$H_{0}$} = 70 km s$^{-1}$Mpc$^{-1}$, \textit{$\Omega_{M}$} = 0.3 
and \textit{$\Omega_{\Lambda}$} = 0.7. In this cosmology, an angular scale of 1 arcsec 
at the distance of the NEP supercluster corresponds to 1.629 kpc. All magnitudes are given 
in the AB system.

\section{THE DATA}

\subsection{IR \& Optical Imaging}

The ANWS was carried out using all available filters of the InfraRed
Camera (IRC). For each of the AKARI cameras, there are three associated 
channels :  NIR (N2, N3, N4), MIR-S (S7, S9W, S11), and MIR-L (L15, L18W, and L24), 
all with a field-of-view covering 10$'$ $\times$ 10$'$. The numbers next to each letter 
represent the central wavelengths in $\mu$m, and the W's for 9 and 18 $\mu$m represent the wider 
bandwidths. The ANWS was completed with 446 pointed observations covering a large area of $\sim$ 
5.4 deg$^{2}$ towards the NEP. Each pointing was done with the `IRC03' Astronomical Observation 
Template (AOT; see AKARI Observer¡'s Manual version
1.2 \footnote{$http://www.ir.isas.jaxa.jp/ASTRO-F/Observation/ObsMan/akobsman12.pdf$}),
with 2 dithered pointings per filter. For detailed descriptions
of the survey strategy, its observational properties and the
reduction of the AKARI IR images we refer the
reader to Matsuhara et al. (2006) and Lee et al. (2009a).
 
The optical survey covers the ANWS field centered at $\alpha$ = 18$^{h}$00$^{m}$00$^{s}$, 
$\delta$ = +66$^{\circ}$36$^{'}$00$^{''}$. The central 2 deg$^{2}$ were covered by the CFHT Megacam 
\textit{u}$^*$, \textit{g}$^\prime$, \textit{r}$^\prime$, \textit{i}$^\prime$, and \textit{z}$^\prime$ 
filters (Hwang et al. 2007). The remaining area, (which includes a small overlap with the CFHT Megacam 
data) is covered with the SNUCAM (Im et al. 2010) on the 1.5m telescope at Maidanak Observatory in 
Uzbekistan using the Bessell \textit{B}, \textit{R}, and \textit{I} filters (Jeon et al. 2010). 
We convert the CFHT Megacam (of ANWS) and SDSS (of A2255) photometry into the photometric system of 
Maidanak (i.e. Bessell $B$ and $R$) using best-fit spectral energy distribution (SED) model colors.
In the following analysis the final uncertainties combine (in quadrature) the original photometric 
errors with the uncertainties derived from the SED fits which are typically $<$ 0.1 magnitudes.

NIR imaging was also carried out using FLAMINGOS on the KPNO 2.1m telescope, which covers 5.2 
and 5.4 deg$^{2}$ in the J and H bands, respectively. Figure 1 shows the coverage of each survey, and the 
mean depth and the FWHM of each band are summarized in Table 1. We did not use the AKARI L24 data in this 
study, due to its insufficient sensitivity. The photometry has been corrected for foreground Galactic 
extinction using the Schlegel et al. (1998) dust maps and Cardelli Milky Way extinction curve 
(Cardelli et al. 1989), assuming $R_{V}$ $=$ 3.1.

The object detection and photometry was done with SExtractor (Bertin \& Arnouts 1996) on the coadded 
images of each individual band. We consider sources as real detections if they have more than five contiguous 
pixels above 3 $\times$ the rms fluctuations of the sky. However it was necessary to match the AKARI 
objects with optical counterparts (FWHM of 0.8$''$ -- 1.4$''$) due to the low resolution (FWHM of 5.5$''$ 
-- 6.6$''$) of AKARI images. Thus, sources that SExtractor cannot separate properly because of 
blending with neighbors in the AKARI IR images are excluded. The photometry was done using  
SExtractor in a single-band mode, and we used MAG-AUTO for the total magnitudes. To check the 
MAG-AUTO values we performed large-aperture photometry for several isolated galaxies in the 
final image. These showed that the difference between MAG-AUTO and MAG-APER were smaller than the 
typical measurement errors (NIR: $\leq$ 5\%, MIR: $\leq$ 20\%). However, MAG-AUTO for sources with 
close neighbors can be easily contaminated by them. To derive fluxes of such objects, we used the
aperture photometry and applied aperture corrections which are derived from the relation between 
second-order moments (SExtractor parameters for measuring the PSF) of the Maidanak $R$ band or 
CFHT $r'$ band and the 11$''$ ($\sim$ 2$\times$FWHM) diameter aperture flux. Figure 2 shows the 
relation between the second-order moments for isolated galaxies in the Maidanak $R$ band image and the
magnitude difference between MAG-AUTO and MAG-APER for $N3$ and $S11$. To determine the aperture 
correction factors, we used the best-fit (red dashed lines) for sample galaxies that are not 
contaminated by nearby sources. 
 

\begin{deluxetable}{ccccccccccccccccccc}
\setlength{\tabcolsep}{2pt}
\tabletypesize{\tiny} 
\tablewidth{0pc} 
\tablecaption{The imaging dataset in the ANWS }

\tablehead{ & \multicolumn{5}{c}{CFHT} & \multicolumn{3}{c}{Maidanak} &    \multicolumn{2}{c}{KPNO} &    \multicolumn{8}{c}{AKARI} \\
               \cline{2-6} \cline{12-19} \\ \\
Covering area & \multicolumn{5}{c}{2 deg$^{2}$}& \multicolumn{3}{c}{4.9 deg$^{2}$} &    \multicolumn{2}{c}{5 deg$^{2}$} &    \multicolumn{8}{c}{5.4 deg$^{2}$} \\ \\
Band  & u$^{*}$ & g' & r' & i' & z'& B & R & I & J & H & N2 & N3 & N4 & S7 & S9W & S11 & L15 & L18W \\ \\
Depth [AB mag.] & 26.0 & 26.1 & 25.6 & 24.7 & 23.7 & 23.2 & 22.0 & 21.2 & 20.5 & 19.3 & 20.9 & 21.1 & 21.1 & 19.5 & 19.3 & 19.0 & 18.6 & 18.8 \\ \\
FWHM  [arcsec]  & 1.13 & 1.05 & 0.93 & 0.84 & 0.79 & 1.4 & 1.2 & 1.1 & 1.4 & 1.3 & 5.5 & 6.0 & 6.0 & 5.9 & 6.6 & 5.9 & 6.2 & 6.2 \\ 
      & (1)  & (2)  & (3)  & (4) & (5) & (6) & (7) & (8) & (9) & (10) & (11) & (12) & (13) & (14) & (15) & (16) & (17) & (18)}

\startdata

\enddata
\begin{flushleft}
Note. $-$ Col. (1)-(5): CFHT Megacam u$^{*}$, g', r', i', and z' from Hwang et al. (2007). 
Col. (6)-(8): Maidanak B, R, and I from Jeon et al. (2010). 
Col. (9)-(10): KPNO Flamingo J and H from Jeon et al. (2011, in preparation). 
Col. (11)-(18): AKARI N2, N3, N4, S7, S9W, S11, L15, and L18W from Lee et al. (2009).  
\end{flushleft}
\normalsize
\end{deluxetable}


\begin{figure}[ht!]
\epsscale{.50}
\plotone{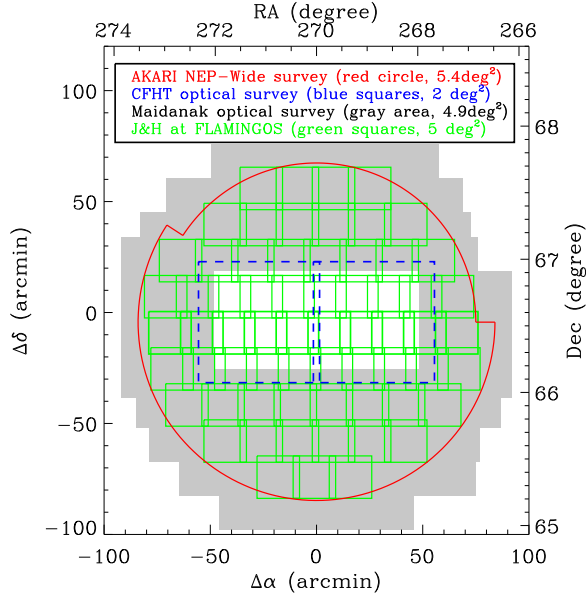}
\caption{ The imaging coverage of the ANWS.
          The center coordinates of the ANWS are at
          ($\alpha$ = 18$^{h}$00$^{m}$00$^{s}$, $\delta$ = +66$^{\circ}$36$^{'}$00$^{''}$).
          North is up and east is to the left.
          The AKARI IR (2--24 $\mu$m) data (5.4 deg$^{2}$), CFHT Megacam 
          \textit{u$^{*}$g$^\prime$r$^\prime$i$^\prime$z$^\prime$} data (2 deg$^{2}$),
          Maidanak \textit{BRI} data (4.9 deg$^{2}$), KPNO \textit{JH} data (5 deg$^{2}$)
          are represented by the red solid circle, blue dashed
          squares, the grey shaded region, and green squares,
          respectively.  \label{fig1}}
\end{figure}


\begin{figure}[ht!]
\epsscale{1}
\plottwo{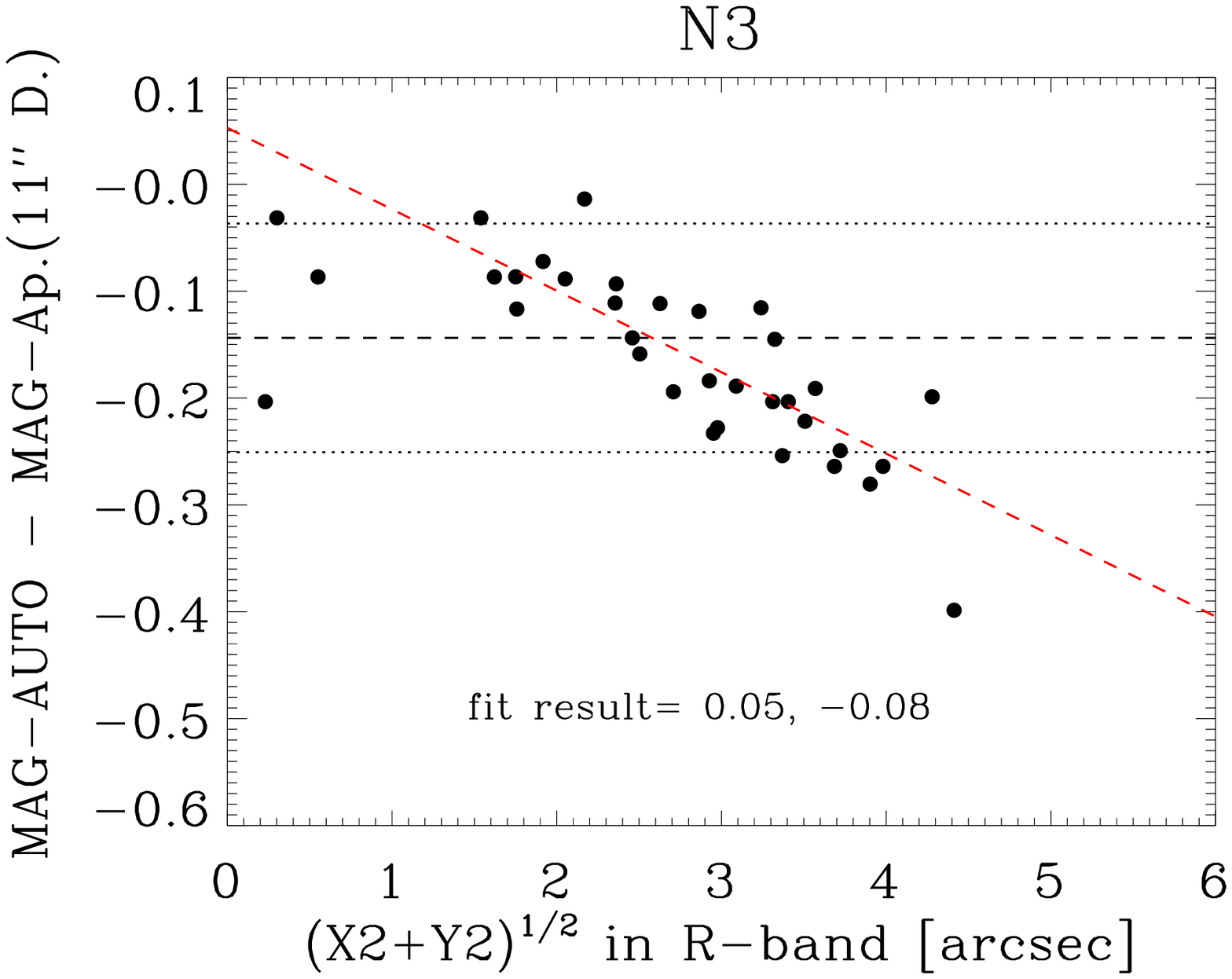}{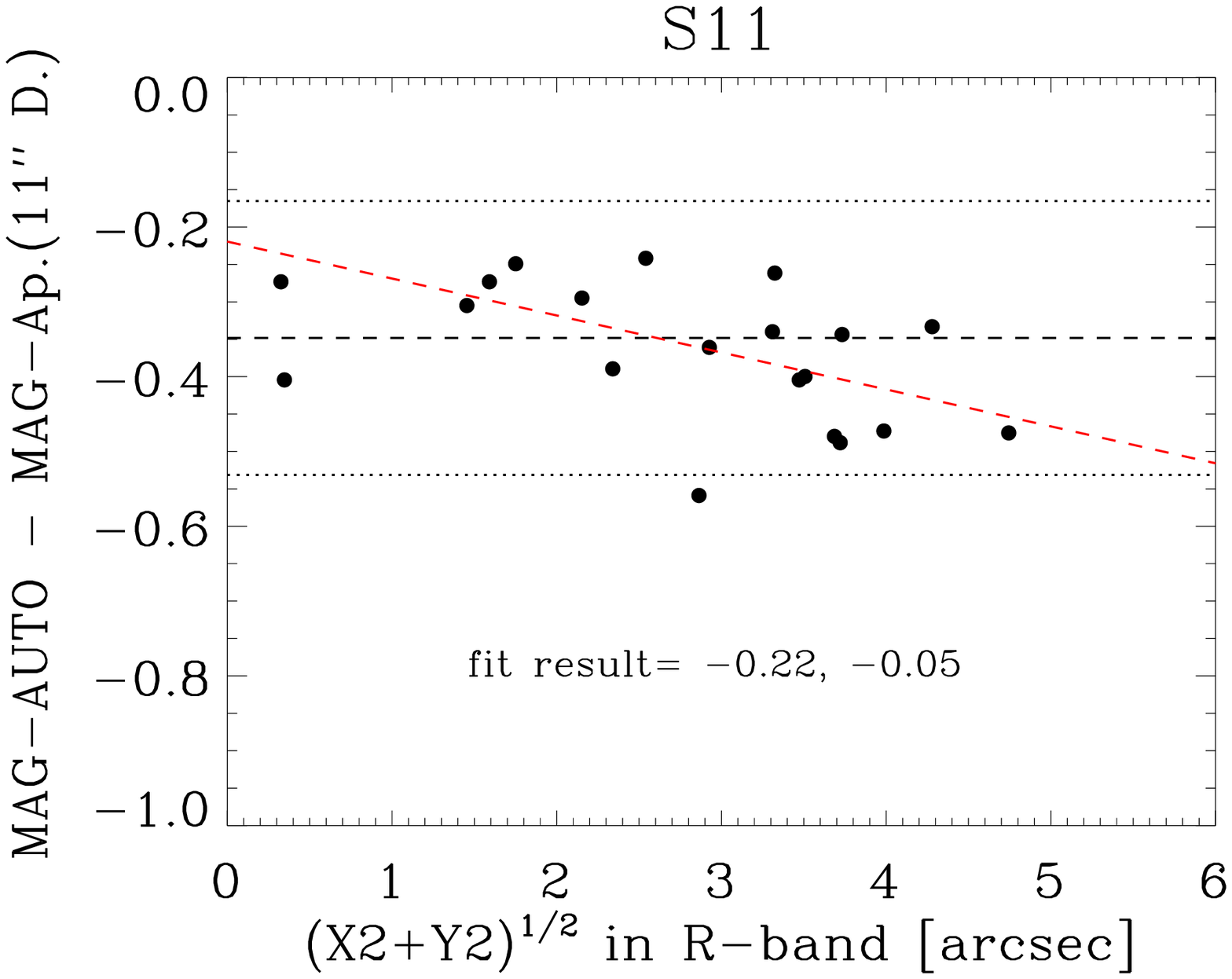}
\caption{ Second-order moments (from SExtractor) in the Maidanak R band image versus magnitude difference 
between MAG-AUTO and MAG-APER with 11$''$ diameter for N3 ($left$) and S11 ($right$). To determine 
aperture correction factors, we used the best-fit (red dashed lines) for galaxies which are not contaminated
by nearby objects. Black dashed and dotted lines show the mean value and the standard deviation of all 
sample galaxies in N3 ($left$) and S11 ($right$), respectively.}
\end{figure}

\subsection{Optical Spectroscopy}

The spectroscopic follow-up of galaxies in the ANWS field used 
MMT/Hectospec and WIYN/Hydra. Based on the optical and IR fluxes, we
selected objects with power-law SEDs as AGN candidates (N2--N4 $>$ 0 and S7--S11 $>$ 0, see Lee et al. 2007) 
and S11-detected objects with 15 $\mu$m flux brighter than 250 $\mu$Jy as SF galaxy candidates. 
Supercluster member candidates (red-sequence galaxies) were selected using the NIR color-magnitude diagram 
(-0.7 $<$ N3--N4 $<$ -0.4), and a brightness requirement in the N2 band (N2 $<$ 18).
This color-cut is adopted as a rough cluster member selection
criterion based on the AKARI study of A2255 galaxies which are at a similar redshift (Shim et al. 2011). 
We also visually inspected the \textit{R} or \textit{r}$^\prime$ band images to exclude 
stars. In summary, we selected as spectroscopic targets galaxies with a wide range of IR fluxes 
to study their IR properties in a variety of local density environments.
However, because of signal to noise limitations, identifying the
correct redshifts of faint absorption-line galaxies is often very
difficult, so that less massive galaxies with absorption line spectra
are under-represented in the sample. We took into account this
incompleteness in our analysis.


\begin{figure}[ht!]
\epsscale{.50}
\plotone{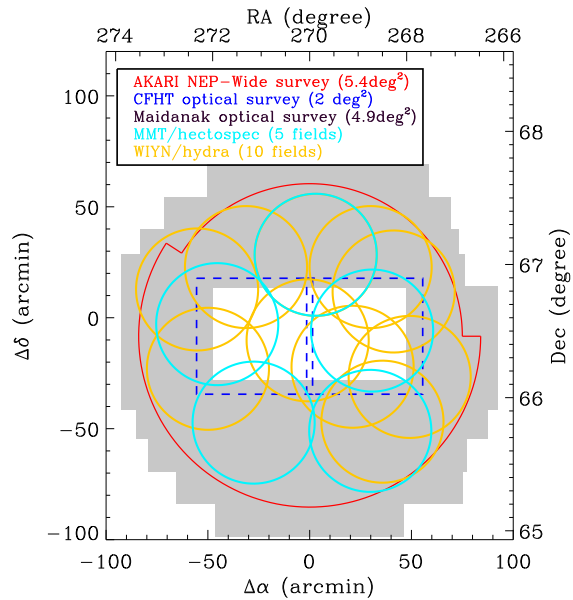}
\caption{ Follow-up spectroscopic observations of the NEP-Wide survey. 
          The five cyan circles indicate MMT/Hectospec fields, 
          and the ten yellow circles indicate WIYN/Hydra fields.
          The representation of the AKARI and optical surveys are
              the same as in Fig.1. \label{fig2}}
\end{figure}

\subsubsection{MMT Hectospec Observations}

Hectospec is a multiobject, moderate-dispersion spectrograph, covering a 1$^{\circ}$ 
diameter field of view at the f/5 focus of the 6.5m MMT (Fabricant et al. 2005). It 
comprises 300 fibers of 1.$^{\prime\prime}$5 diameter, covering a wavelength range of 
3500--9000 \AA{} with 6 \AA{} FWHM resolution.

The Hectospec observations are taken in queue mode, so that the field is targeted when 
optimally placed on the sky. A total of five configurations were obtained between 2008 
May and November. A log of the observations is shown in Table 2, which contains the field
identification, the J2000.0 coordinates, total exposure time and date.

The Hectospec observations used a catalog of galaxies detected in the AKARI-IR bands 
(high priority objects), complemented by other galaxies with R $\leq$ 22 in the region, 
but that were undetected by AKARI. In addition to galaxies, this catalog also contains candidate F 
stars, selected from the photometry, which are used to flux-calibrate the spectra. The 
assignment of objects to fibers is done using 
XFITFIBS\footnote{http://www.harvard.edu/john/xfitfibs}, which takes into account the 
number of configurations (5 in the case of the ANWS), the object priorities and 
number of sky positions. Typically 250 fibers per setup were assigned to NEP objects, 
40 to random sky positions and up to 6 fibers placed on the candidate F stars. The 
spectroscopic reduction used the {\it{HSRED}} package of $IDL$ scripts written by 
R. Cool\footnote{http://www.astro.princeton.edu/~rcool/hsred/hsred/reductions.html},
which is based on the $IDL$ pipeline developed for the reduction of SDSS spectra. 
{\it{HSRED}} does the standard reduction by correcting for bias, flatfields, 
illumination (if twilight flats were taken), performing wavelength calibration 
(from HeNeAr lamps), sky-subtraction and extracting one-dimensional spectra. The 
flux-calibration is done by combining the 1-dimensional F-star spectra with the 
multiband photometry (to obtain the spectro-photometric zero-point) and Kurucz 
stellar models (to rectify the spectra). Redshifts for the wavelength- and 
flux-calibrated spectra are obtained from the cross-correlation with a series of
galaxy, QSO and stellar template spectra. All Hectospec redshifts were individually 
validated and assigned a quality code ranging from 1 to 4, as used for the DEEP2 
survey (e.g., Newman et al. in preparation; Willmer et al. 2006). Only qualities of 4 or 3
are used in the analyses,  meaning that the probability of the redshift being correct is  
greater than 95\% and 90\% respectively.

\begin{deluxetable}{ccccc}
\tabletypesize{\scriptsize} 
\tablewidth{0pc} 
\tablecaption{MMT Hectospec fields \label{tab-numden}}

\tablehead{
Field &  R.A. (J2000)    &  DEC. (J2000)   &  t$_{exp}$ (minutes) & Observation Date (UTC) }

\startdata

nep-hecto-1 & 17 54 50.66 & +66 38 47.48 & 100 & 2008 May 03 \\
nep-hecto-2 & 18 04 29.63 & +65 53 32.20 & 80 & 2008 June 02 \\
nep-hecto-3 & 17 55 09.27 & +65 49 28.81 & 80 & 2008 September 03 \\
nep-hecto-4 & 17 59 29.51 & +67 16 03.83 & 80 & 2008 November 17 \\
nep-hecto-5 & 18 07 39.24 & +66 38 56.78 & 80 & 2008 November 20 \\

\enddata

\end{deluxetable}

\subsubsection{WIYN Hydra Observations}

We also obtained optical spectra with the Hydra multiobject spectrograph on the WIYN 3.5m telescope 
at Kitt Peak National Observatory. We used 98 red fibers of 2$''$ diameter feeding the bench spectrograph 
with a 316 lines mm$^{-1}$ grating, yielding a dispersion 2.64 of \AA{ }pixel$^{-1}$. The wavelength 
range is 4500--9000 \AA, but the spectrum quality is low beyond 8000 \AA{} due to the strong sky 
emission lines. The field of view of Hydra is approximately 1$^{\circ}$ and we observed ten fields 
over the ANWS field (Fig. 3).

Target assignment in each configuration was done using the WHYDRA software. For each configuration, 
10--15 fibers were assigned to blank sky positions, and 3--6 fibers were assigned to spectrophotometry 
standard stars. Excluding broken fibers, fibers assigned to blank sky positions, and fibers assigned 
to standard stars, we obtained spectra of 60--70 targets in each configuration. In Figure 3, the 
locations of the Hydra configurations are shown by ten yellow circles, and details of the observations
are summarized in Table 3. Depending on the observing conditions, the exposure time for 
each field varied from 3$\times$20 minutes to 5$\times$20 minutes.

We used IRAF to reduce the spectra. First, we performed the
pre-processing which includes the corrections for overscan, bias, dark
and flat, and trimming the image. A flatfield image was created by averaging dome flats 
taken before and after the observations. We removed cosmic rays using L.A.Cosmic (van Dokkum 2001).
After the pre-processing, we extracted one-dimensional spectra using the Hydra reduction package 
DOHYDRA (F. Valdes 1995)\footnote{Guide to the HYDRA Reduction Task DOHYDRA, available at 
$http://iraf.net/irafdocs/dohydra.pdf$}. We extracted one-dimensional spectra from all apertures and 
did wavelength calibration with a Cu--Ar comparison lamp. The master sky spectrum produced by coadding 
sky spectra of blank skies, was subtracted with DOHYDRA. Finally, the extracted one-dimensional spectral 
images were combined using IRAF task $scombine$ to improve the final signal-to-noise ratio.

\begin{deluxetable}{ccccc}
\tabletypesize{\scriptsize} 
\tablewidth{0pc} 
\tablecaption{WIYN Hydra fields \label{tab-numden}}

\tablehead{
Field &  R.A. (J2000)    &  DEC. (J2000)   &  t$_{exp}$ (minutes) & Observation Date (UTC) }

\startdata

NEP00 & 18 00 10.031 & +66 34 00.00 & 80 & 2008 June 27 \\
NEP01 & 17 56 29.855 & +66 21 00.00 & 80 & 2008 June 27 \\
NEP02 & 17 54 09.796 & +65 54 00.00 & 100 & 2008 June 27 \\
NEP03 & 17 51 51.806 & +66 16 00.00 & 80 & 2008 June 28 \\
NEP04 & 17 52 58.195 & +66 58 00.00 & 80 & 2008 June 28 \\
NEP05 & 17 54 51.441 & +67 10 00.00 & 60 & 2008 June 29 \\
NEP06 & 17 59 28.904 & +67 16 00.00 & 60 & 2008 June 30 \\
NEP07 & 18 05 22.322 & +67 10 00.00 & 60 & 2008 June 30 \\
NEP08 & 18 09 25.763 & +66 59 00.00 & 60 & 2008 June 30 \\
NEP09 & 18 08 18.026 & +66 20 00.00 & 60 & 2008 June 30 \\

\enddata

\end{deluxetable}

\subsubsection{Redshift Identification}

Redshifts were determined by identifying high signal-to-noise emission lines
and/or multiple absorption lines. Three individuals (J. Ko, M. Im, and H. Shim) independently 
determined redshifts for all objects. Then, each of them flagged objects according 
to their spectral features. We flagged the objects with at least two distinct spectral 
features as those with a secure redshift. All individuals generally agree on the secure 
redshifts, but faint or distant galaxies can be ambiguous due to the weak line features. 
These objects were not used in the analyses.

To verify our redshift determination, we ran $XCSAO$ in the $RVSAO$ package, which computes radial 
velocities by cross-correlating spectra against templates of known redshift (Kurtz et al. 1992). 
In this test, we used 61 supercluster member galaxies flagged as having secure redshifts with 
Hectospec galaxy templates, and found that the difference of radial velocities for all samples 
is 119$\pm$123 km s$^{-1}$, consistent with no difference in radial velocities measured from both 
methods and telescopes.

We were able to successfully determine secure redshifts for 1026 and 400 of
1195 Hectospec objects and 600 Hydra objects, respectively. Figure 4 shows a sample of WIYN 
spectra with lines used to determine redshifts, and the redshift distribution of all objects 
with secure redshifts in the ANWS is shown in Figure 5. 
To this sample we added 241 galaxies for which redshifts are available in the NASA Extragalactic 
Database (NED).
An additional 18 redshifts come from long slit spectra obtained by Matthew Malkan using the Kast 
spectrograph at the Lick Observatory 3-m telescope.

\begin{figure}[ht!]
\epsscale{0.9}
\plotone{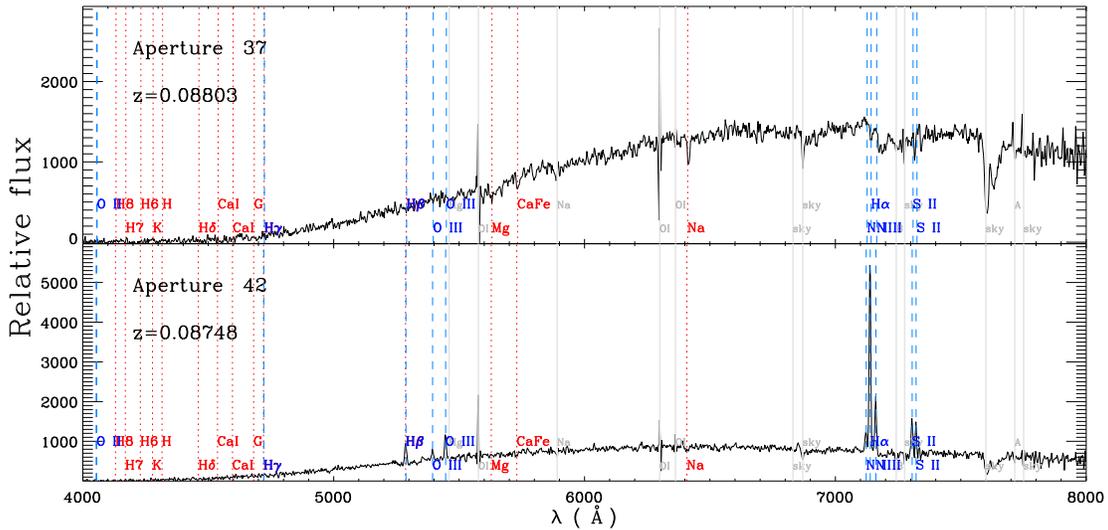}
\caption{Sample of WIYN/Hydra spectra with sky (grey), absorption (red), and emission (blue) lines used
for redshift determination. \label{fig4}}
\end{figure}

\begin{figure}[ht!]
\epsscale{0.8}
\plotone{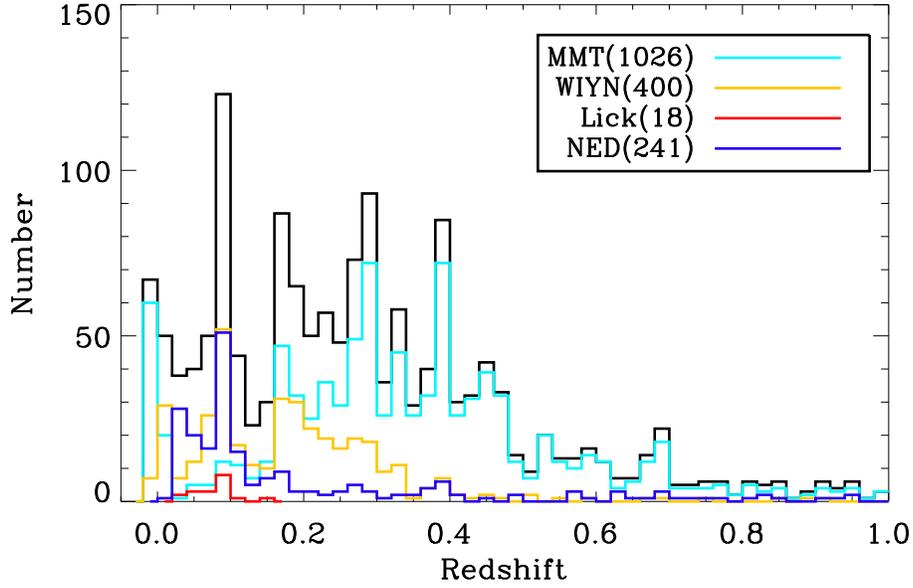}
\caption{Redshift distribution of galaxies with secure redshifts in the ANWS. 
         The black line shows the redshift distribution of the total spectroscopic sample. \label{fig5}}
\end{figure}

In Figure 6, we plot the spectroscopic completeness as a function of the observed $N3$ magnitude 
and ($B-R$) color for extended sources. Here we use $N3$ band and ($B-R$) color because we will 
adopt the absolute $N3$ band as a rough stellar mass indicator and ($B-R$) color to separate red 
sequence galaxies from blue cloud galaxies. The vertical line represents our N3 magnitude cut (N3 $<$ 19) 
used in this study.

\begin{figure}[ht!]
\epsscale{0.8}
\plotone{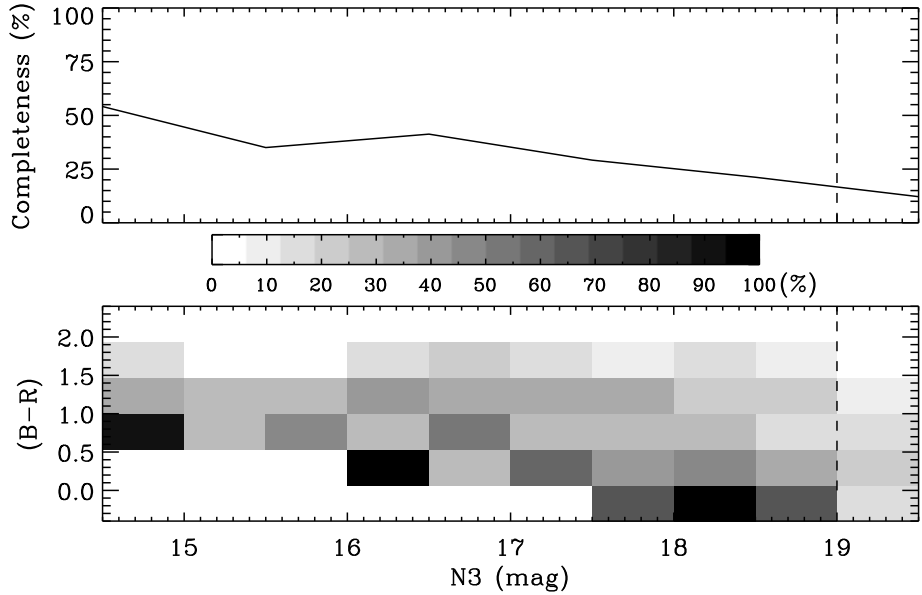}
\caption{Spectroscopic completeness as a function of the N3 apparent
  magnitude (top panel) and 
observed ($B-R$) color and N3 apparent magnitude (bottom panel) of galaxies in the ANWS.
Vertical dashed lines indicate the N3 magnitude cut used in this study.  \label{fig6}}
\end{figure}
\clearpage

\subsection{Supercluster Member Selection}

The sample of NEP supercluster members used in this paper consists of the spectroscopic sample 
of galaxies at 0.07$<$z$<$0.1 in the field of ANWS and of galaxies in A2255 (see Fig. 7). 

After removing duplications, we have a total of 150 secure spectroscopic objects in the redshift 
interval 0.07$<$z$<$0.1 in the ANWS. Using the optical positions, we matched the spectroscopic 
data with catalogs derived from our AKARI IR data, KPNO NIR data ($J$ and $H$), and GALEX UV data 
where available. The matching radius was chosen to be 5.0 arcseconds. All matched objects were then 
visually validated by examining postage stamp images for all bands.
  
The A2255 spectroscopic and photometric data are drawn from Shim et al. (2011). We used 313 objects 
covered by the GALEX UV data. The redshift identification is nearly complete for galaxies with 
r $<$ 18.9 mag. The limiting optical magnitude places a limit in the derived infrared 
luminosity, which will be taken into account in following analysis. As for the ANWS data, the 
redshift range for NEP supercluster member galaxies is 0.07$<$z$<$0.1. The detection limits for 
observed bands are 25, 30, 65, 80, 150, and 400 $\mu$Jy in the N3, N4, S7, S11, L15, and L24 bands, 
respectively (Shim et al. 2011). The depth of the A2255 data is comparable to that of the ANWS data 
within the measurement errors. 

\begin{figure}[ht!]
\epsscale{0.9}
\plotone{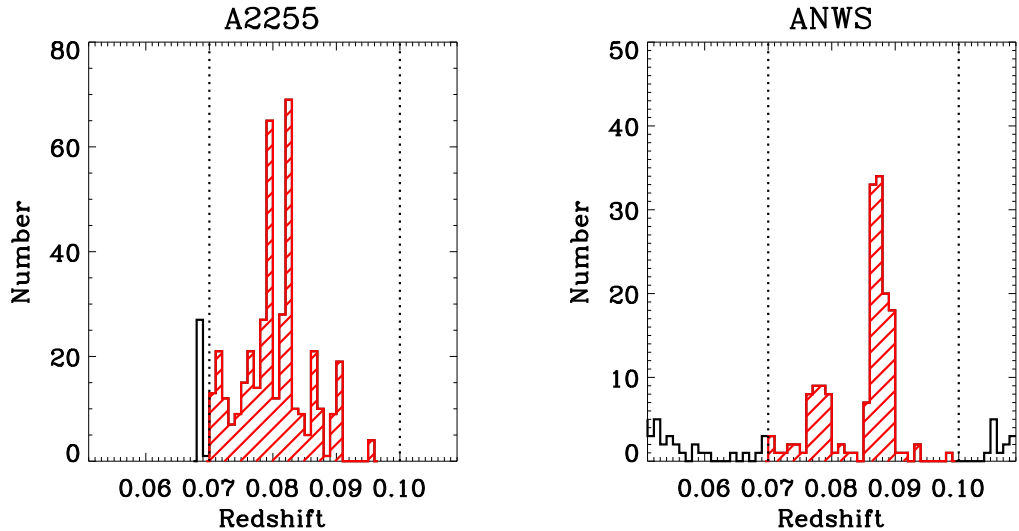}
\caption{The redshift distribution of galaxies in A2255 ($left$) and the ANWS ($right$).
         The vertical dotted lines indicate the redshift range of the NEP supercluster
         (0.07$<$z$<$0.1).}
\end{figure}
\clearpage

\section{PHYSICAL PROPERTIES}

In this section, we describe the method used to derive the galaxy stellar masses, SFR, and 
local galaxy density.

\subsection{Spectral Energy Distributions}

We derive the stellar mass, the mean stellar age, the reddening parameter, and the SFR for each object 
through SED fitting, which uses a standard $\chi^{2}$ minimization procedure with various templates. 
We used two different SED libraries -- the Bruzual \& Charlot (2003; BC03) spectral synthesis models 
to estimate the stellar mass and age, and the IR templates of SF galaxies from Chary \& Elbaz (2001; CE01) 
to determine the IR luminosity.
When performing the fits for SF galaxies, we only use the IR data in the fits, as the empirical IR 
templates do not include the diverse range of ages, metallicities and star formation histories that are
modeled by the BC03 library. Figure 8 shows examples of the observed UV-optical-MIR SEDs of galaxies in 
the ANWS, and the best-fit SEDs from each model are overplotted.

\begin{figure}[ht!]
\epsscale{.9}
\plotone{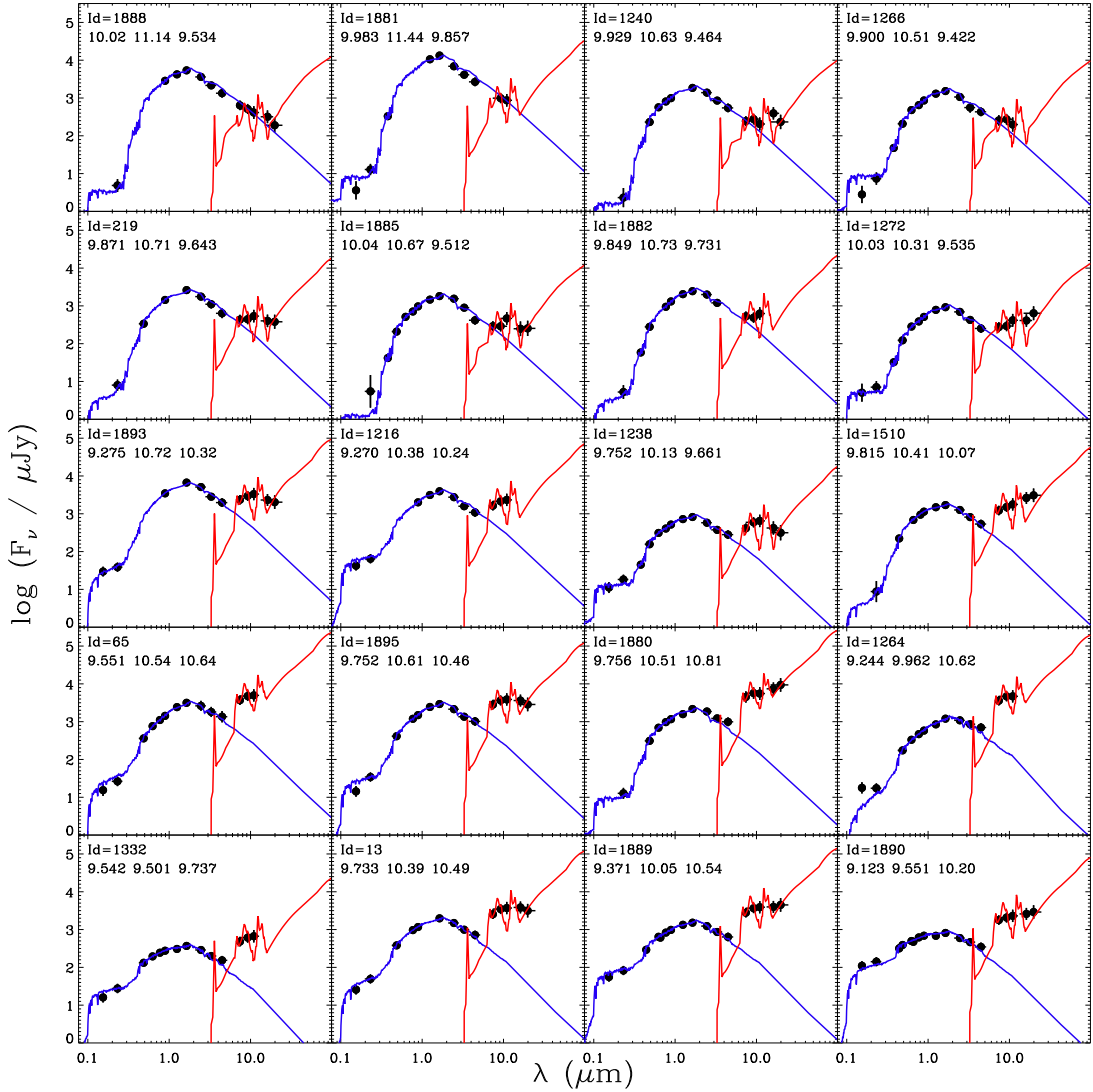}
\caption{Examples of the SEDs of galaxies in the ANWS. 
         Filled circles indicate observed data points from GALEX UV to AKARI MIR.
         All objects are detected in S11, except in cases where objects
         were not detected in the AKARI L15 and L18 bands
         (Id: 1881, 1266, 1882, 1216, 65, 1264, and 1332), in which
         cases no points are shown. 
         Overplotted lines represent the best-fit SEDs: 
         the blue solid lines indicate SEDs from Bruzual \& Charlot (2003) and the red solid lines 
         indicate infrared galaxy templates of Chary \& Elbaz (2001).
         Each SED fit lists the object id on top-left, stellar age (years), stellar mass ($M_{\odot}$), 
         and total $L_{IR}$ ($L_{\odot}$) in unit of dex, from left to right, respectively.
         Each row shows a set of four objects that are representative of our classification 
         scheme (described in Table 4):
         weak-MXG, intermediate-MXG, weak-SFG, dusty-SFG, and blue-SFG, from top to bottom.}
\end{figure}
\clearpage

Stellar masses were estimated using stellar population
synthesis models (e.g., Bell et al. 2003;  Ilbert et al. 2010). During
the SED-fitting process, the redshift was 
kept fixed, and all bands from the
NUV to NIR (NUV -- N4) were used to fit a purely stellar SED. The SED templates 
were generated with BC03 models assuming a Chabrier (2003) initial mass function (IMF) and
an exponentially declining star formation history SFR $\propto$ $e^{-t/\tau}$ ($\tau$ between 
0.1 Gyr to 30 Gyr).
The SEDs were generated for a grid of 44 ages (0.1 Gyr to 13.5 Gyr) and three different 
metallicities (0.02, 0.008, and 0.004 Z$_{\odot}$), and dust
extinction was added using the formula of Calzetti 
et al. (2000) for $E_{B-V}$ between 0 to 0.5. Figure 9 shows the stellar masses computed 
with BC03 versus the $N3$ absolute magnitude of galaxies in the
ANWS, which shows a good correlation. This suggests that $N3$ 
is a rough indicator of the stellar mass. As M$_{N3}$ = -19 corresponds to
a stellar mass of $\sim$ log $M_{*}$ = 9.1 $M_{\odot}$, the stellar mass range of our sample is roughly
log $M_{*}$ = 9.1 -- 11.5 $M_{\odot}$. The histogram of the $N3$ absolute magnitude of galaxies is 
shown in the upper panel of Figure 10, where the top axis shows stellar masses corresponding to the 
N3 absolute magnitude (Fig. 9). 

\begin{figure}[ht!]
\epsscale{0.6}
\plotone{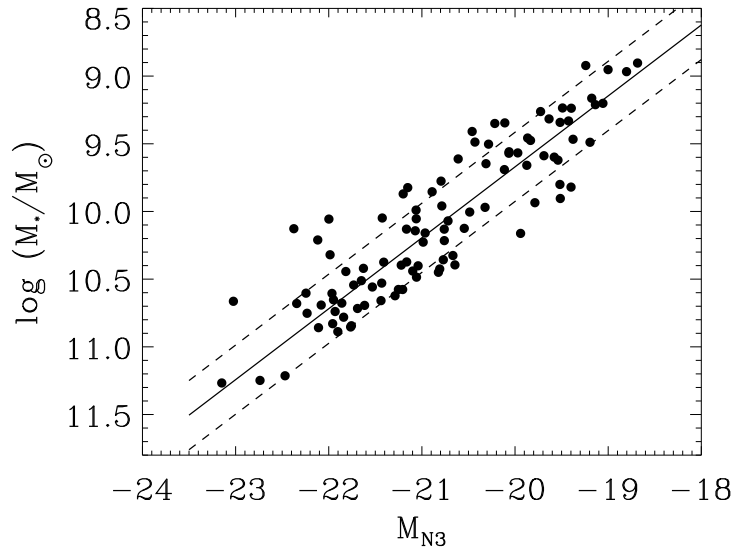}
\caption{The stellar masses computed with BC03 vs. the N3 absolute magnitude of galaxies in the ANWS.
         Here we used 102 galaxies for which 5 or more bands were used in the SED fits.
         The solid line indicates the best-fit (log $M_{*}$/M$_{\odot}$ = --0.524 $\times$ M$_{N3}$ -- 0.810), 
         and the two dashed lines represent the standard deviation of residuals to this fit ($\sigma$ = 0.255).}
\end{figure}

\begin{figure}[ht!]
\epsscale{0.8}
\plotone{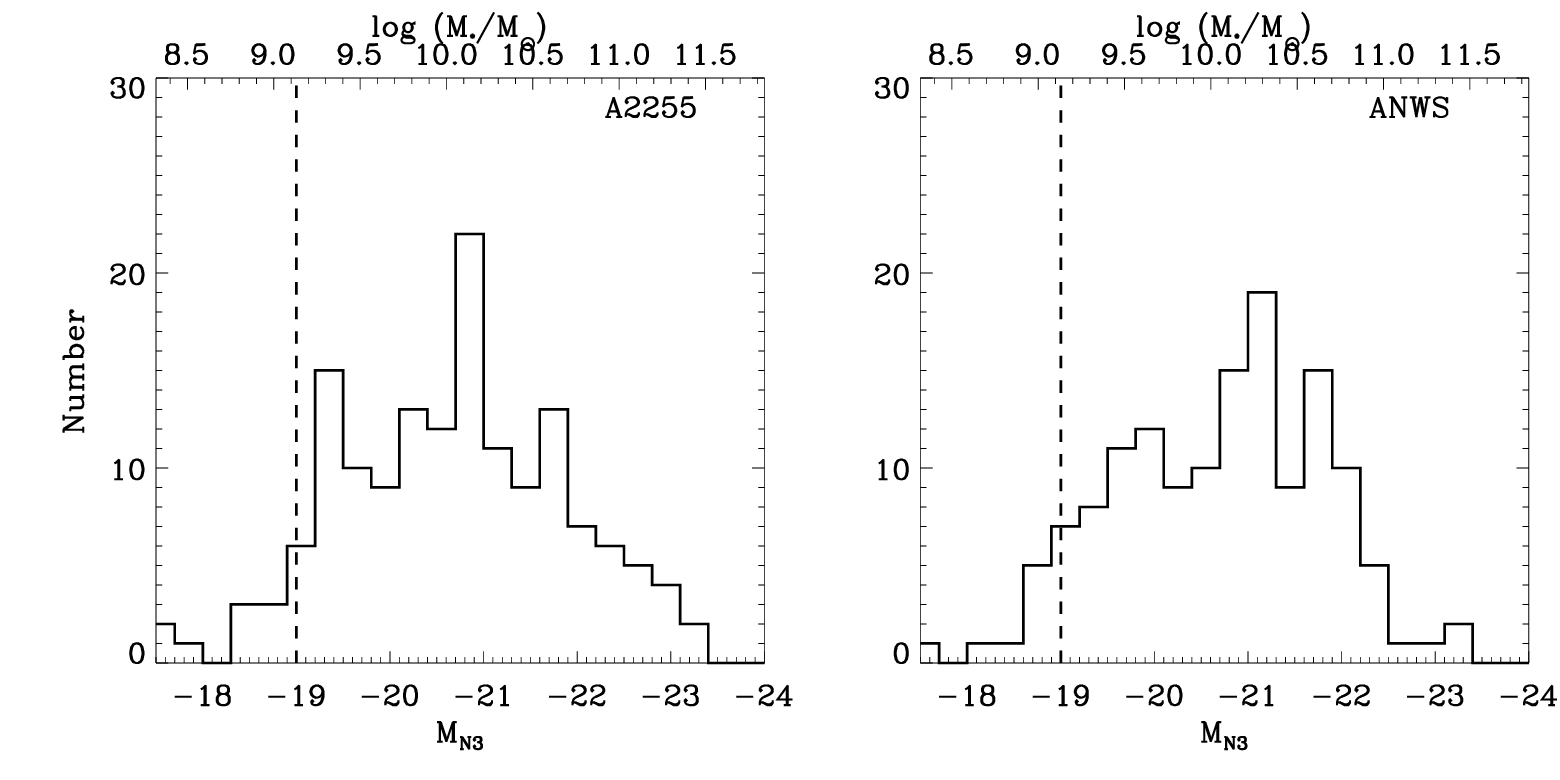}
\plotone{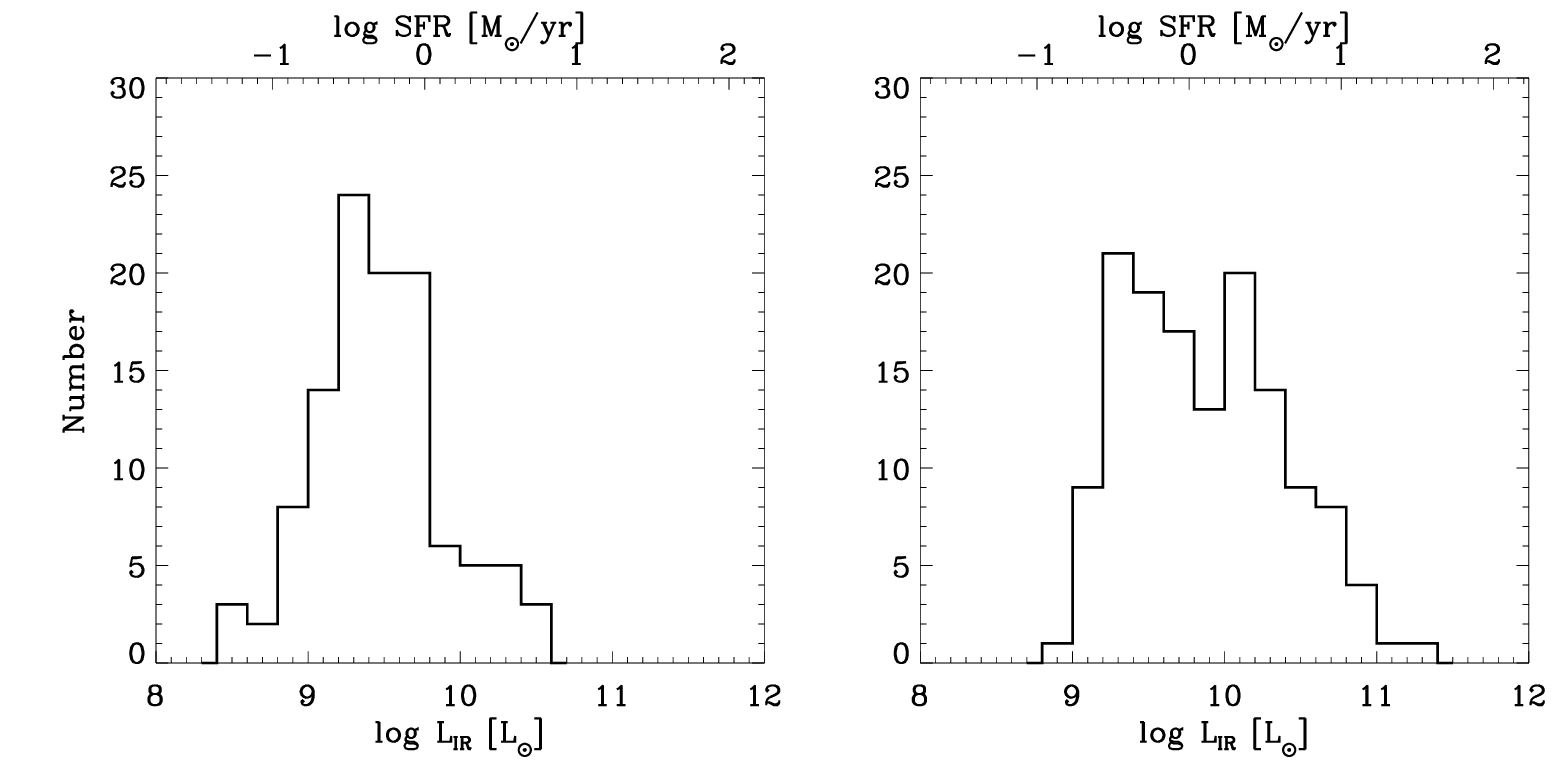}
\caption{\textit{Upper}: Histogram of the N3 absolute magnitude of galaxies in A2255 ($left$) 
         and in the ANWS ($right$). The top axis shows stellar masses calculated from
         the linear relation in Fig. 9. Vertical lines are our N3 magnitude cut.
         \textit{Lower}: Distribution of derived total IR luminosity $L_{IR}$ (8--1000) for S11 
          detected galaxies in A2255 ($left$) and the ANWS ($right$).
          The Top axis corresponds to SFR, using the  Kennicutt (1998)
          relation. }
\end{figure}

To derive the $L_{IR}$ for each galaxy, we use the IR templates of CE01. Templates are shifted to 
each galaxy's redshift and then matched to all the available bands longer than 7$\mu$m (i.e. 
the AKARI S7, S9W, S11, L15, and L18W bands). We first subtract the stellar contribution 
(i.e. the best-fit BC03 template to the UV--NIR data) from the observed IR data, and then 
fit them with CE01 templates to estimate the total IR luminosity at 8-1000 $\mu$m ($L_{IR}$). 
We convert $L_{IR}$ into SFR using Kennicutt (1998) relation: 
SFR $(M_{\odot} yr^{-1})$ = 1.72 $\times$ $10^{-10}$ $L_{IR}/L_{\odot}$. The lower panel of 
Figure 10 shows the distribution of derived total IR luminosity $L_{IR}$ for S11 detected galaxies. 
We only find two Luminous IR Galaxies (LIRG; log L$_{IR}/L_{\odot}$ $>$ 11) candidates in the ANWS field.
Conservatively, we estimate a SFR limit of 0.1 $M_{\odot} yr^{-1}$ in A2255 and 
0.2 $M_{\odot} yr^{-1}$ in the ANWS. There is a likely contribution of AGN to the 
IR luminosity. However, the contamination in our sample is very low if we assume 
many AGNs have power-law IR SEDs (e.g., Lee et al. 2007). Among the supercluster 
member galaxies in the ANWS, there is only one X-ray point source (detected by ROSAT), 
which is also a LIRG. Thus, AGN contamination is expected to be negligible in our results.

\clearpage

\subsection{Galaxy Local Density}

Although our dataset does not cover the entire NEP supercluster, it covers a wide range of 
environments, including a rich galaxy cluster (A2255), three galaxy
groups, and low density regions in the supercluster outskirts.

To characterize the galaxy properties as a function of the environment, we adopt 
a local surface density estimator $\Sigma$, which is the surface number density of galaxies 
within a projected distance of 0.5 Mpc and within a relative velocity of 1000 km s$^{-1}$ for 
each galaxy in the sample. The velocity cut is adopted to exclude foreground and background galaxies. 
Our method of measuring galaxy environment (i.e. the fixed aperture environment measure)
is found to be the best estimate of galaxy environmental density when the virial radius
of the host halo is difficult to measure according to Haas et al. (2011), so we chose 
this as our density parameter for discussion.
As a cautionary measure, we compared several other environment indicators against this 
measure, using galaxies in A2255. Two other indicators are tested, one is the surface 
number density of galaxies within a projected distance of 1.0 Mpc and 
within a relative velocity of 1000 km s$^{-1}$ ($\Sigma_{1.0Mpc}$), and another is 
the surface number density of galaxies within to the 5th-nearest neighbor and  
within a relative velocity of 1000 km s$^{-1}$ ($\Sigma_{5th-nearest}$). 
In the right panel of Figure 11, we show the correlation between the local density 
adopted in this work ($\Sigma_{0.5Mpc}$) and other environmental parameters. 
The Spearman rank correlation coefficient is also printed, indicating other choice 
of environmental parameters correlate well with our choice.
Indeed, we performed the analysis on our main results using these different environment 
parameters, and find that the results are not affected by a choice of the environmental parameter.
To account for the spectroscopic incompleteness of the NEP-Wide survey, the number density is computed 
by weighing each galaxy by the inverse of completeness corresponding to its $N3$ apparent 
magnitude and ($B-R$) color in Figure 6. In the left panel of Figure 11 shows the distribution of the local 
surface density for galaxies in A2255 (dotted line) and in the ANWS (dashed line).
The peak density in the ANWS area corresponds to the lowest density region of A2255, 
i.e. the infall region of A2255. We refer to this as an intermediate local density 
in the following analysis. The spatial distribution of all supercluster member 
galaxies in A2255 and in the ANWS field is overplotted on their number density map in Figure 12.
 
\begin{figure}[ht!]
\epsscale{1}
\plottwo{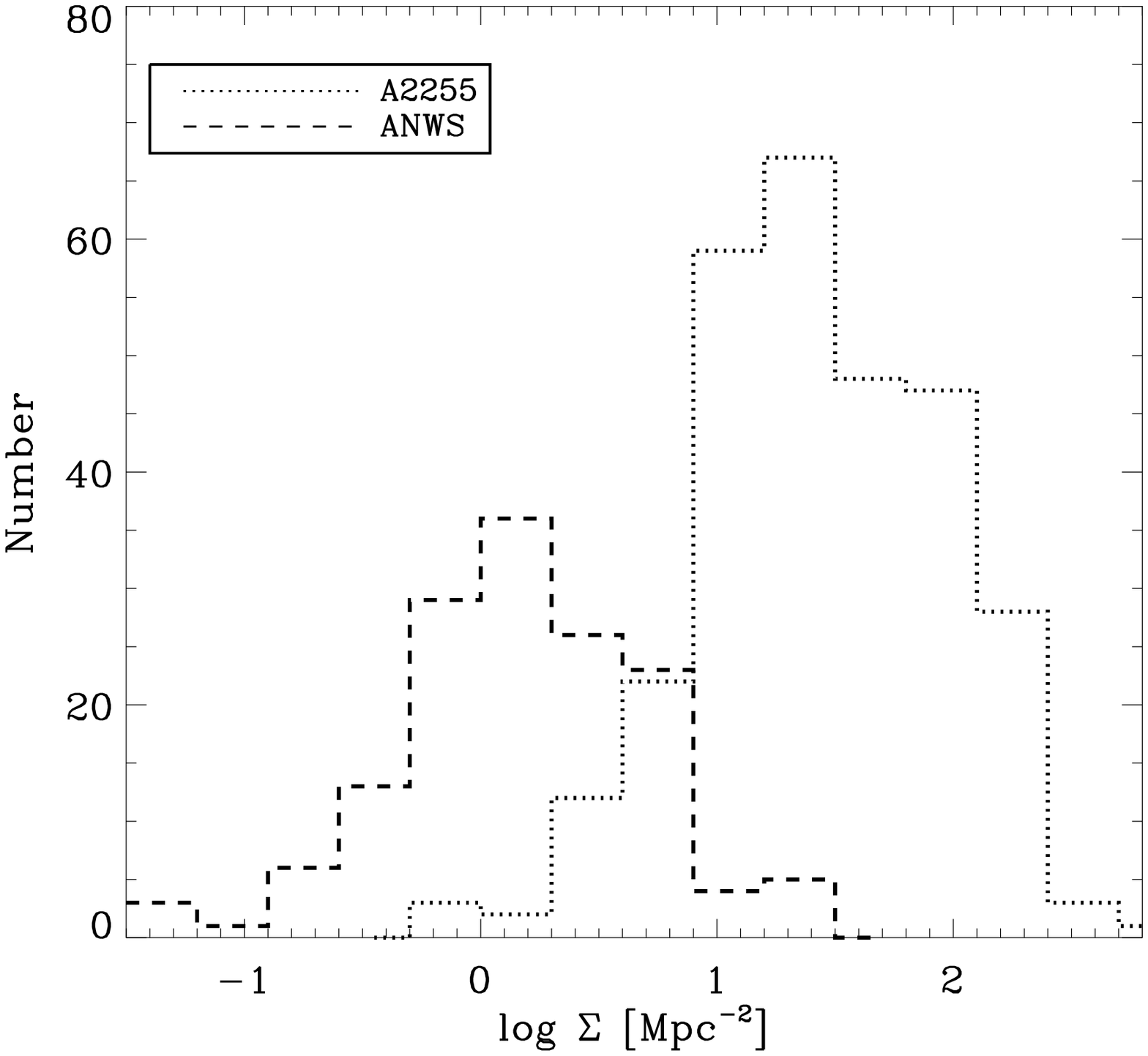}{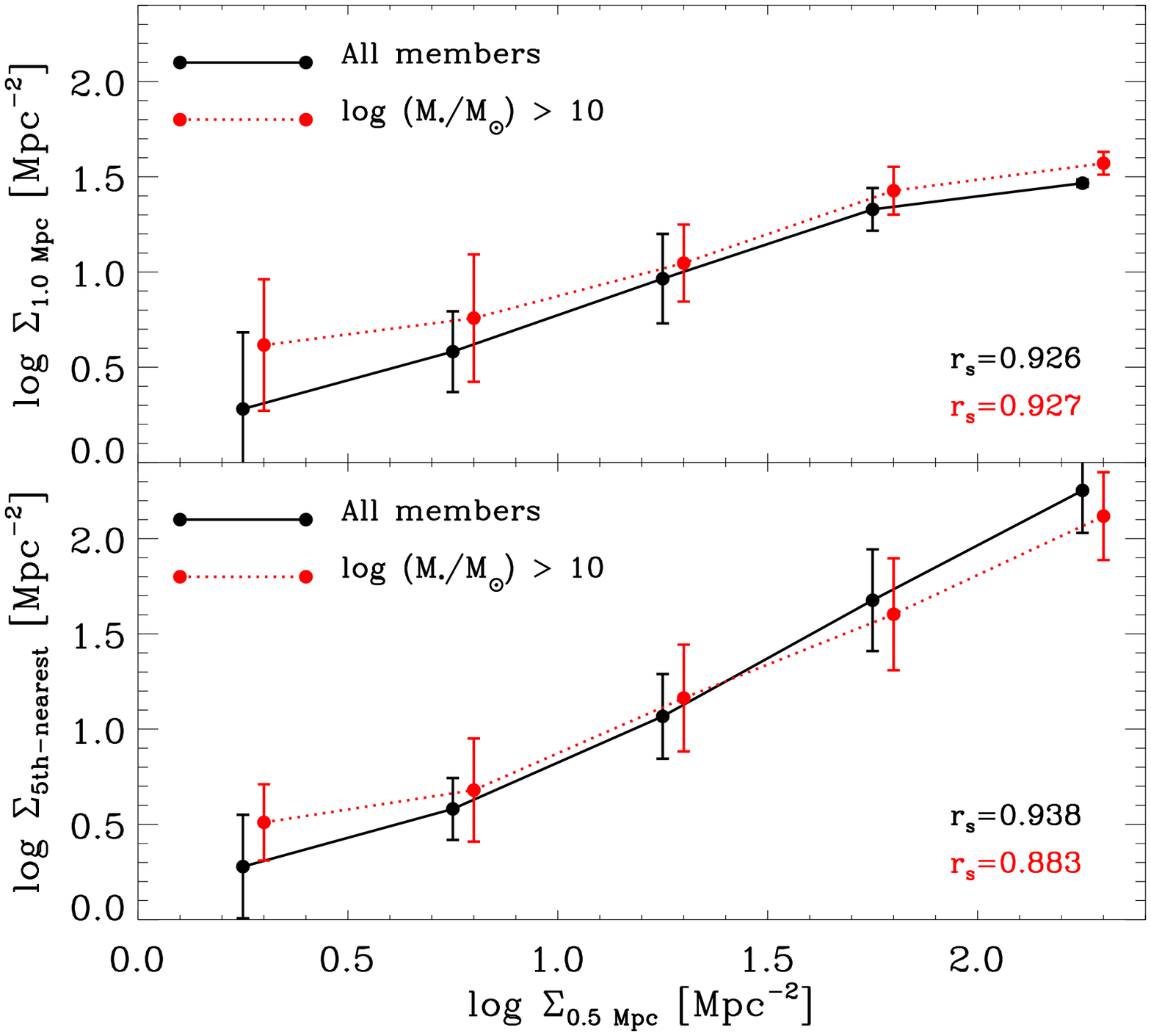}
\caption{\textit{Left}: Distribution of the local density $\Sigma$ for galaxies in A2255 (dotted line) 
         and in the ANWS (dashed line).
         \textit{Right}: Compasiron between the local density $\Sigma_{0.5Mpc}$ adopted in this work, 
                          which is the number density within a projected distance of 0.5 Mpc, and
                          $\Sigma_{1.0Mpc}$ within a projected distance of 1.0 Mpc ($upper$) and
                          $\Sigma_{5th-nearest}$ within a projected distance to 5th-nearest neighbor ($lower$).
                          The neighbors were identified among all members (black-solid lines) and
                          massive (log $M_{*}$/M$_{\odot}$ $>$ 10) ones (red-dotted lines) 
                          within $\Delta v$ $<$ 1000 km s$^{-1}$. The Spearman rank correlation coefficient
                          $r_{s}$ between local densities are printed in the bottom-right of each panel.}
\end{figure}

\begin{figure}[ht!]
\epsscale{1}
\plotone{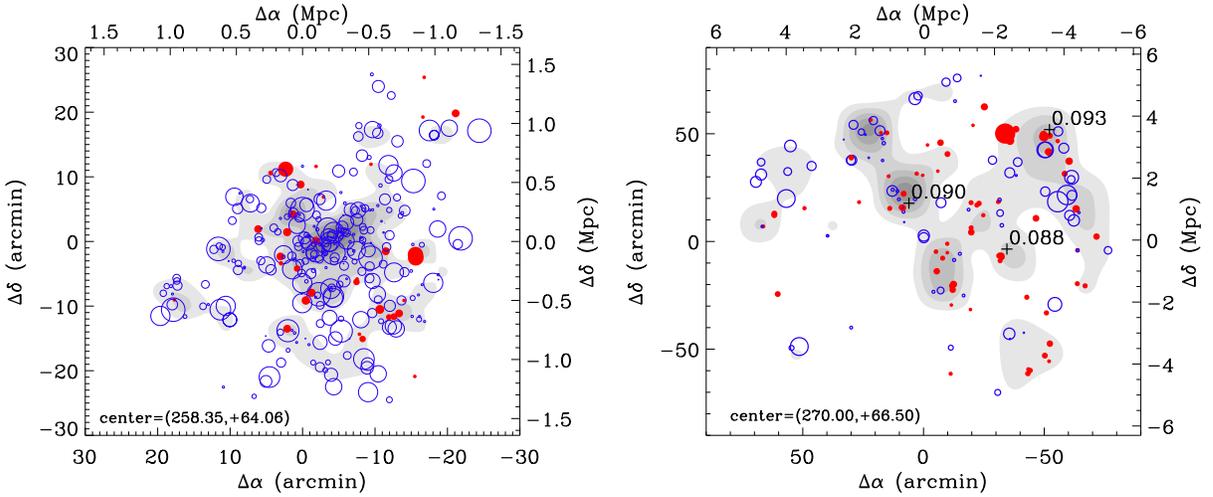}
\caption{ The spatial distribution of galaxies in A2255 ($left$) and the ANWS ($right$),
          on the smoothed galaxy number density map.  
          Red or blue circles represent galaxies with redshifts above
          or below z=0.087, respectively. 
          The symbol size is proportional to the redshift deviation from the NEP supercluster
          mean redshift (z=0.087).}
\end{figure}
\clearpage

\section{GALAXY CLASSIFICATIONS: MIR View of Galaxies}

\subsection{Motivation}

The optical color-magnitude relation (CMR) is commonly used to separate red-sequence (hereafter red) 
galaxies from blue-cloud (hereafter blue) galaxies. Blue optical colors can mean that galaxies are 
actively star-forming. While the presence of an AGN can make a galaxy blue (e.g., Obric et al. 2006; 
Choi et al. 2009), SF is in general the cause of blue galaxy colors. 
The red colors of galaxies can be due to a dominant population of passively evolving old stars 
(even though some galaxies with predominantly old stars have low-level SF, 
e.g., Trager et al. 2000). Galaxies can also be red in the optical/NUV if their 
light is extinguished by interstellar dust. Thus we need to differentiate such 
causes of red galaxy colors.

We show the CM diagram of A2255 galaxies (the left side of Fig. 
13) compared to galaxies in the ANWS (the right side of Fig. 13) in three colors: 
$B-R$ (panels a, d),  $NUV-R$ (panels b, e), and $N3-S11$ (panels c, f). The absolute 
$N3$ magnitude (horizontal axis) is used as a rough measure of the stellar mass (see Fig. 9). 
To determine the rest frame $M_{N3}$ absolute magnitude, we apply only the luminosity distance 
of each galaxy (i.e. no correction for peculiar motions or bandpass shift are included). In the 
following analysis, we only consider galaxies brighter than $M_{N3}$ = -19 (corresponding 
$N3$ $\approx$ 19) due to the relatively shallow $S11$ detection limit. Specifically, this 
cut allows us to construct an unbiased sample of galaxies with $N3-S11 >$ 0 (see next section). 

In Fig. 13(a,d), we show the $B-R$ optical CMR for galaxies brighter than $M_{N3}$ = $-$19 in A2255. 
The red-sequence is defined using a linear fit to the observed $B-R$ versus $M_{N3}$, shown as a dashed line, 
by rejecting outliers iteratively based on the bi-weight calculation.
The standard deviation of residuals to the fit is 0.07 mag ($\sigma$), implying a tight optical red-sequence.
The horizontal solid line indicates the color cut we adopted to separate red galaxies 
(redward of the solid line) from blue galaxies (blueward of the solid line). 
The CMR was moved to a bluer color by $\Delta(B-R)=0.21 (\sim3\sigma)$ to define 
the color cut. The same color criterion is applied to the galaxies in the ANWS.

Fig. 13(b,e) shows the $NUV-R$ CM diagram for the same sample as in panels (a,d). The NUV CM diagram shows 
a scatter an order-of-magnitude larger than that in the optical data,
indicating that  a number of red galaxies have been 
forming stars (e.g., Yi et al. 2005). Since the NUV flux is much more sensitive to young stellar populations 
than the optical flux, it can be a good tracer of recent star-formation (within $\sim$ 1.5 Gyr). 
However, while the NUV flux is also sensitive to dust extinction, the
MIR flux is not. 
The wide dispersion in the $N3-S11$ colors for red galaxies in the MIR
CM diagram (Fig. 13(c,f)), suggests that these galaxies present a
variety of MIR properties.

We focus in particular on the MIR. The choice of the $S11$ band has several advantages over other MIR bands.
First, the $S11$ flux correlates with the Polycyclic Aromatic Hydrocarbons (PAHs) emission features at 11.3 
and 12.7 $\mu$m, which may be related to current star formation. Because of the PAH features, MIR emission 
(especially around 11 $\mu$m) correlates well with the total IR luminosity (Spinoglio et al. 1995), which can 
be converted into SFR (Chary \& Elbaz 2001). Second, the MIR emission may also contain a contribution from 
the envelopes of evolved stars, showing broad silicate emission features around 10 $\mu$m 
(e.g., Bressan et al. 2006) or/and unusual PAH festures (e.g., Vega et al. 2010). In particular, this MIR 
emission from dust surrounding AGB stars is also sensitive to stellar ages, because it declines with 
time (e.g., Piovan et al. 2003; Temi et al. 2005). 
Thus the MIR emission may be tracing not only the current SF, but also past SFA. 
For the model SEDs in Figure 13(c), we used Single Stellar Population 
(SSP) models including the dust emission from circumstellar dust around AGB stars (Piovan et al. 2003; 
hereafter P03). As has been recognized, the optical CMR can be well-described with a single age 
model assuming a metallicity gradient (the dashed line in Fig. 13(a); e.g., Kodama et al. 1997). 
The same model fits the $N3-S11$ versus $N3$ CMR at three different stellar ages (1, 5 , and 12 Gyr), 
shown by dotted lines. The horizontal solid line represents the SSP model without AGB dust. However, 
any single age model fails to reproduce the dispersion in the $N3-S11$ colors (see Ko et al. 2009),
which suggests either the existence of younger stellar populations or of some other mechanism.

We divide the red galaxies into several subsamples with different SFA
by using the $N3-S11$ colors.  
We note that NIR--MIR color can be considered as an indicator of the specific SFR (SFR/$M_{*}$; SSFR) 
of SF galaxies, and also of the luminosity-weighted mean stellar age in passively evolving objects.

\begin{figure}[ht!]
\epsscale{1.0}
\plottwo{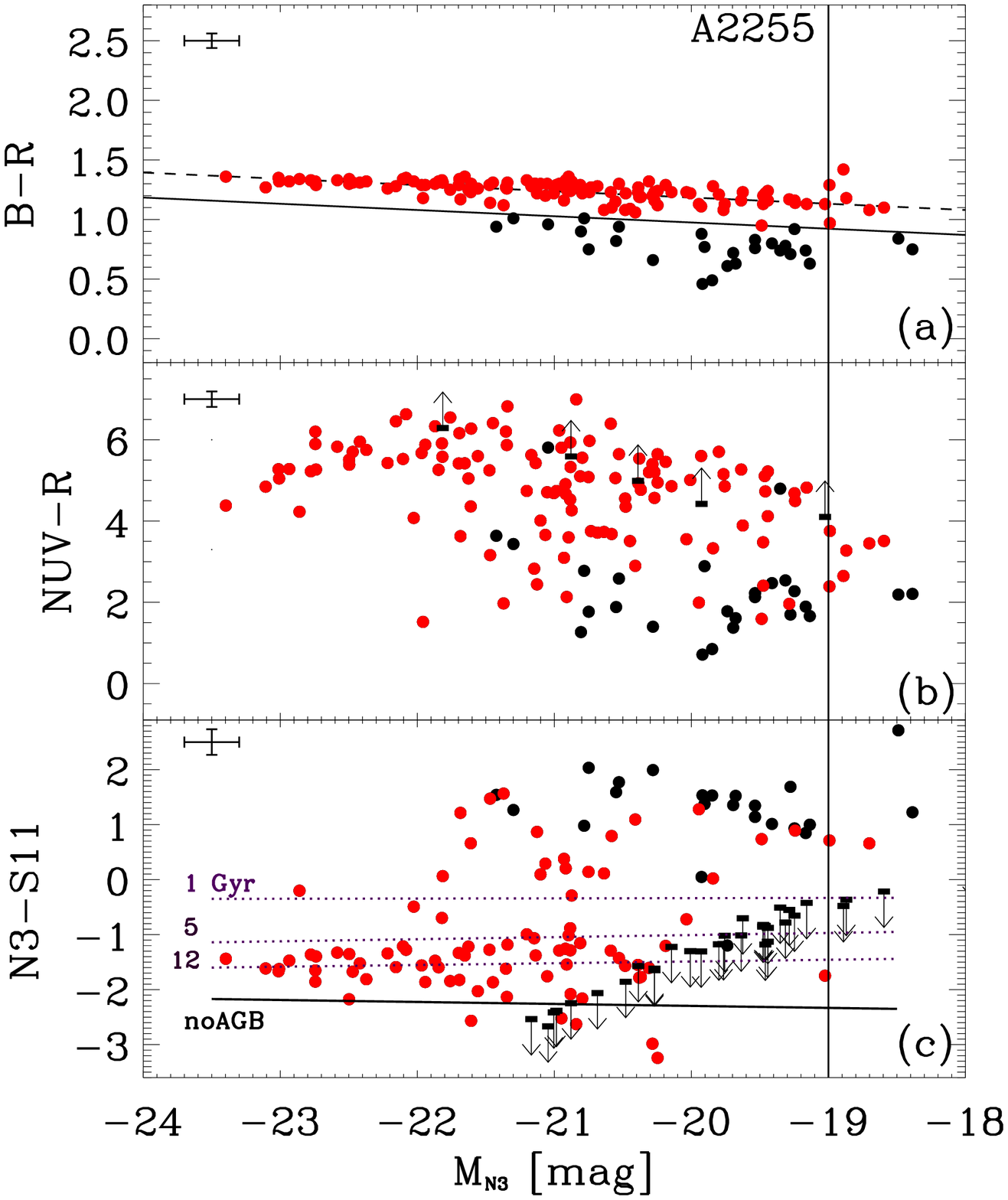}{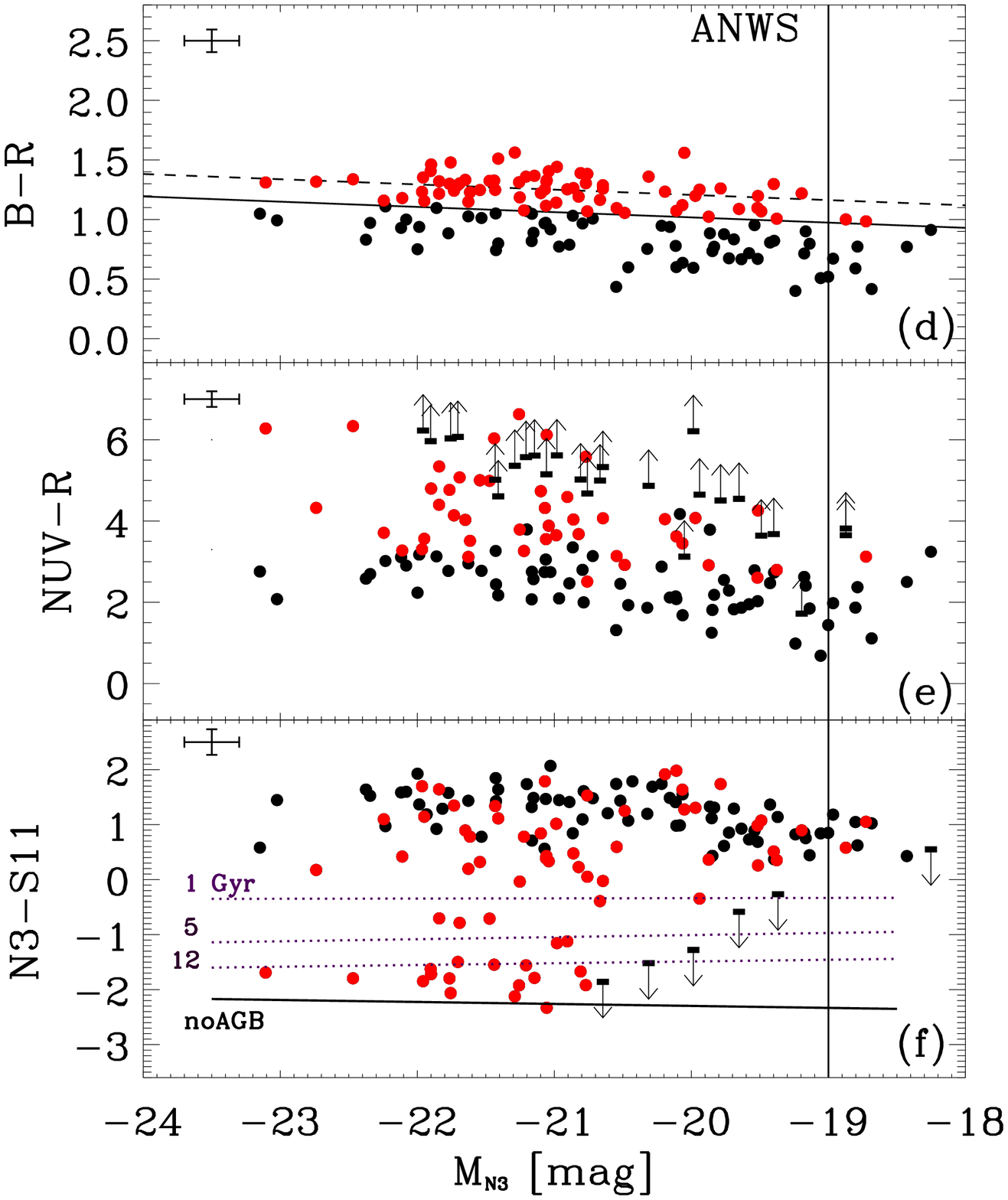} 
\caption{\textit{Top}: the $B-R$ vs. absolute $N3$ color-magnitude (CM) diagram for galaxies brighter 
                       than $M_{N3}$ = $-$19 in A2255 ($left$) and in the ANWS ($right$). 
                       The CMR is shown by the dashed line and the horizontal solid line indicates the
                       color cut adopted to classify red galaxies (redward of the solid line). 
                       The CMR in the ANWS is normalized to the red sequence of A2255.
                       The color deviation from the CMR is defined as $\Delta(B-R)=0.21 (\sim3\sigma)$, 
                       where $\sigma$ is the standard deviation of residuals to CMR fit.
                       The cross in the upper left corner indicates the typical errors.
      \textit{Middle}: the $NUV-R$ vs. absolute $N3$ CM diagram for galaxies in A2255
                       ($left$) and in the ANWS ($right$).
                       The $GALEX$ Nearby Galaxy Survey $NUV$ detection limits are shown for undetected galaxies 
                       (arrows).
      \textit{Bottom}: the $N3-S11$ vs. absolute $N3$ CM diagram for galaxies in A2255
                       ($left$) and in the ANWS ($right$).
                       The dotted lines indicate the CMR calculated from 
                       the P03 AGB model SSPs, assuming a metallicity sequence at three different 
                       stellar ages (1, 5 , and 12 Gyr), respectively. 
                       The horizontal solid line represents the P03 model SSPs without AGB dust. 
                       The $AKARI$ $S11$ detection limits are given for undetected samples in $S11$ 
                       (arrows). The vertical line indicates the magnitude cut ($N3$ $\approx$ 19 
                       corresponding to M$_{N3}$ = -19 at the distance of the NEP supercluster).}
\end{figure}

\clearpage

\subsection{MIR Classification}

Using the SWIRE ($Spitzer$ Wide-area InfraRed Extragalactic survey) templates of Polleta et al. (2007),
we find that early-type galaxies (e.g., ellipticals, S0, and Sa) have $N3-S11$ $<$ 0, while 
SF, late-type galaxies have $N3-S11$ $>$ 0 (see Fig. 14). These templates have been successfully 
used in a number of works for all galaxy types at a range of redshifts (e.g., Adami et al. 2008; 
Ilbert et al. 2009).
Among the SWIRE templates, we use 16 templates including 3 ellipticals (with 2, 5, and 13 Gyr), 
7 spirals, and 6 starbursts covering the wavelength range between 0.1 and 1000 $\mu$m. These are 
generated with the GRASIL code (Silva et al. 1998) including dusty envelopes of AGB stars following
the prescription by Bressan et al. (1998).
In particular, in the MIR spectral region between 5 and 12 $\mu$m, the spiral 
and starburst templates encompass the variety of observed MIR spectra (Polleta et al. 2007). 
Also shown as a comparison are SSP templates of P03 incorporating AGB dust with a metallicity (Z=0.02) 
and three stellar ages (2, 5, and 13 Gyr). 
The P03 model SEDs show somewhat redder $N3-S11$ colors than the GRASIL model due to 
difference in how dust emission from evolved stars is predicted, but they are broadly 
in agreement that $N3-S11$ color as the mean stellar age decreases.
For comparison, we also plot SSPs from different libraries (BC03, CB07, and Ma05) with 
a Salpeter (1955) IMF and a fixed metallicity (Z=0.02), but without the inclusion of 
the MIR dust emission prescription. CB07 is a new version of BC03 including the new 
stellar evolution prescription of Marigo \& Girardi (2007) for the thermally-pulsing 
(TP) AGB evolution, and Ma05 is also including the TP-AGB phase of stellar evolution 
(Maraston 2005) differently from previous models. However, these two SSPs (CB07 and Ma05) 
do not include the circumstelalr dust emissions. 
These three models without AGB dust cannot produce the MIR-excess colors of $N3-S11 > -2$ 
of red and early-type galaxies, although the circumstellar dust formation and the evolution of
AGB stars are still known to be uncertain in detail. 
From two model SEDs including AGB dust we can set a threshold in MIR color ($N3-S11$) 
to divide galaxies into MIR-red galaxies ($N3-S11$ $>$ 0) and MIR-blue galaxies 
($N3-S11$ $<$ 0). Also, $N3-S11 > -1$ divides relatively young (2-5 Gyr old) passive 
ellipticals and old ellipticals. In summary, the MIR-red galaxies are dominated by SF, late-types 
in our sample because of little contamination by AGNs, while most of MIR-blue galaxies 
are early-types showing a wide range of MIR-weighted mean stellar ages.

\begin{figure}[ht!]
\epsscale{1.0}
\plottwo{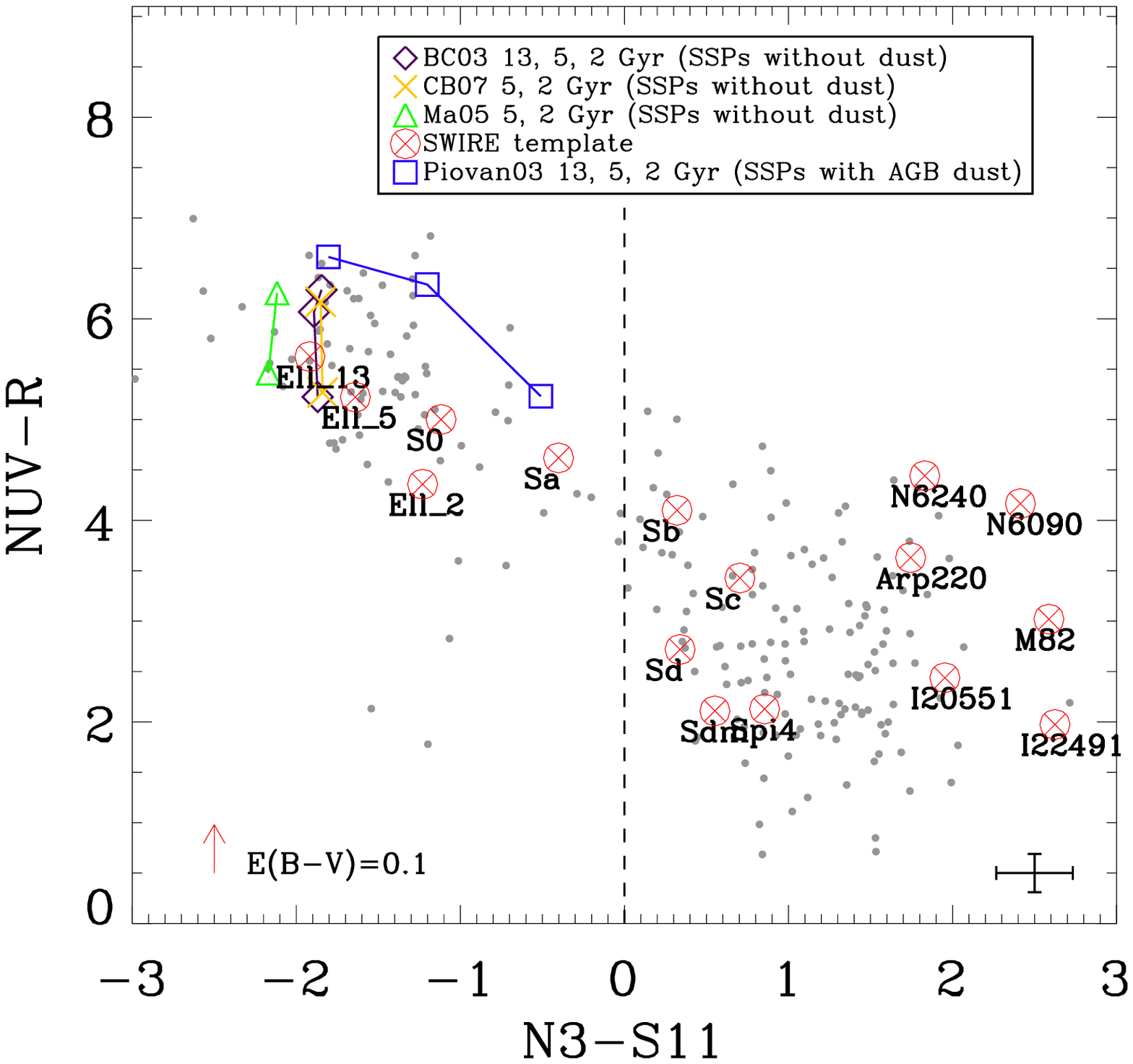}{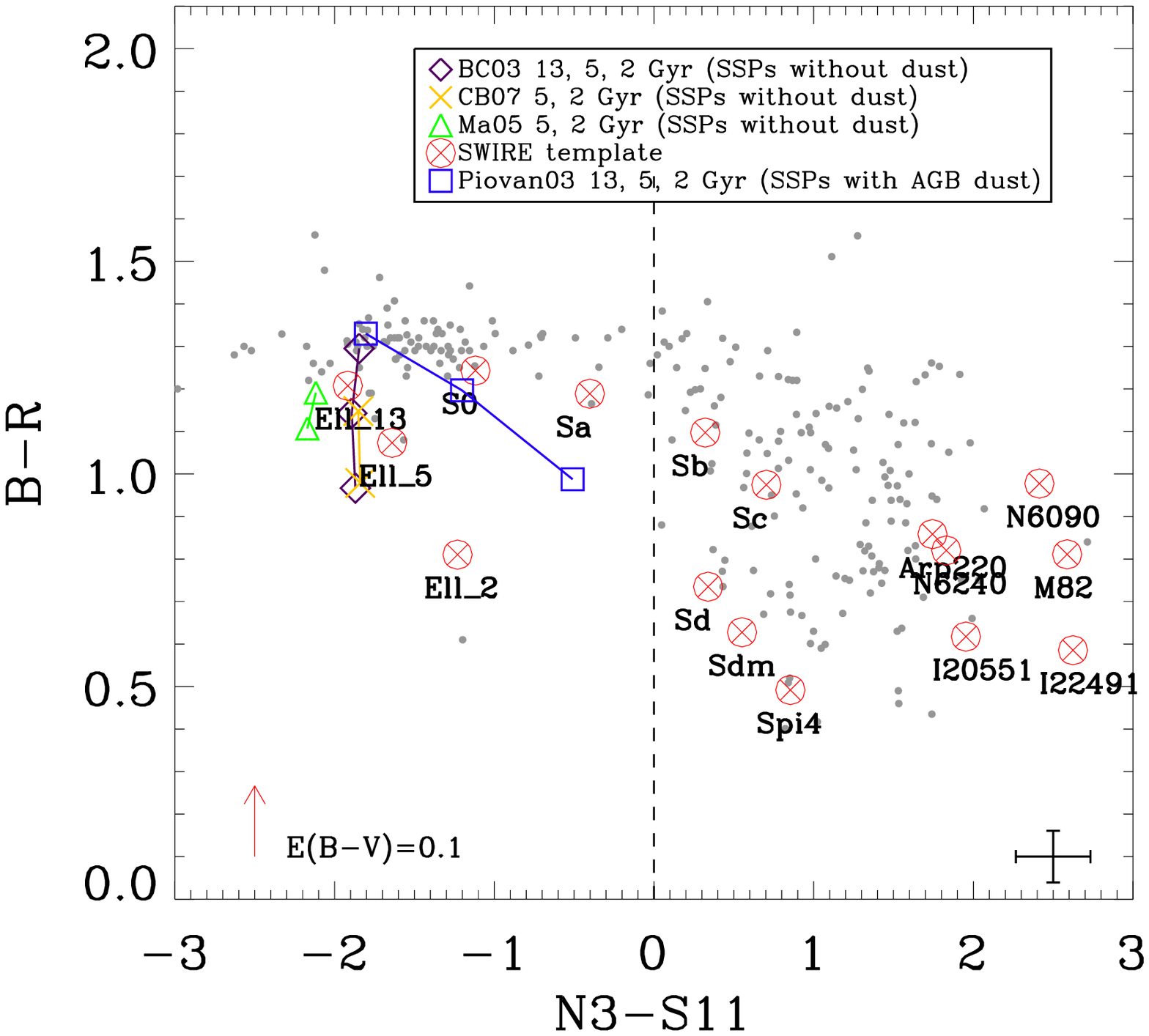}
\caption{The rest-frame $N3-S11$ vs. $NUV-R$ ($left$) and $N3-S11$ vs. $B-R$ ($right$) color-color  
         distribution of galaxies detected in all four bands, compared with 
         SWIRE templates of Polletta et al. (2007) including  3 ellipticals 
         (2, 5, 13 Gyr), 7 spirals (S0, Sa, Sb, Sc, Sd, Sdm, Spi4), and 6 starbursts 
         (M82, Arp220, N6090, N6240, I20551, I22491). Also plotted are
         SSP templates from different libraries (BC03, CB07, Ma05, and Piovan03) 
         with a Salpeter (1995) IMF and a solar metallicity (Z=0.02) for the same ages as SWIRE ellipticals. 
         Templates with $N3-S11 > 0$ are spirals or starbursts, while early types have $N3-S11$~$<$~0.
         The arrow shows the mean value of a reddening vector of E(B-V) = 0.1 using
         the Calzetti et al. (2001) extinction law, and the cross in the bottom right 
         corner indicates the typical errors.}
\end{figure}

Figure 15 shows the $N3-S11$ color versus SSFR derived from the SED fits. This reveals that the 
$N3-S11$ color correlates well with SSFR, indicating that the $N3-S11$ color probes
different levels of SFA. Specifically, we find that our MIR color cut ($N3-S11$ = 0) 
is comparable to log (SSFR) $\sim$ -10.7, i.e. galaxies with log (SSFR) $<$ -10.7 (which 
corresponds to a SFR of 0.2 $M_{\odot}yr^{-1}$ at a stellar mass of $10^{10} M_{\odot}$) 
can be considered as passively evolving galaxies. This SSFR cut was
adopted by Gallazzi et al. (2009) to separate 
SF galaxies from quiescent ones. However, in the case of MIR-blue galaxies, the 
derived SSFR may not have any physical significance as a measure of the current 
SFA, but may indicate a wide dispersion of mean stellar ages among those passive galaxies if the 
P03 model SEDs are adopted.

\begin{figure}[ht!]
\epsscale{1}
\plotone{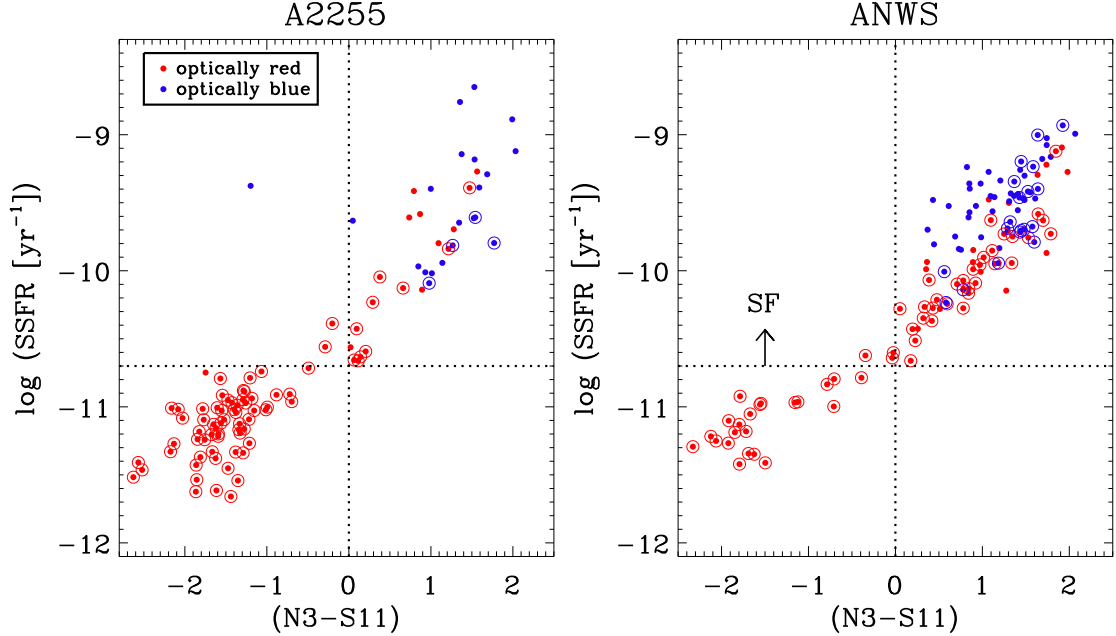}
\caption{SSFRs derived from the SED fits and $N3-S11$ colors of
         galaxies detected in the $N3$ and $S11$
         bands. The filled red and blue circles are optically red and blue galaxies classified by the optical 
         CMR (Fig. 13). The open circle indicates massive galaxies (log $M_{*}$/M$_{\odot}$ $>$ 10). 
         The vertical and horizontal dotted lines indicate our
         NIR$-$MIR color ($N3-S11$ = 0) and 
         SSFR (log(SSFR) = $-$10.7) cuts respectively, that were
         adopted to separate SF galaxies from quiescent ones following Gallazzi et al. (2009).}

\end{figure}

Now we further divide optically red galaxies into four classes depending on their $N3-S11$ colors. 
We made the morphological classification of galaxies in the ANWS field
sample using the CFHT $r'$ band images of 
0.$''$187 pixel$^{-1}$: early types are bulge-dominated with good symmetry (0 and 1 in 
Figs 16-20), while late types are disk-dominated with asymmetric internal structure 
(2, 3, and 4 in Fig. 16-20). The dominant morphological type and its fraction are shown in Table 4. 

First, we classify optically red galaxies with $N3-S11$ $<$ $-$1 (weak MIR-excess galaxies, 
hereafter weak-MXG). The SEDs of these galaxies indicate that they have passively evolving, 
old stellar populations with a mean stellar age greater than 2-5 Gyr.
We note that the absolute age should not be taken too seriously considering that the mean 
stellar ages for a given $N3-S11$ color varies depending on the model used. 
They may have a small amount of MIR-excess at 11 $\mu$m, but that can be well understood 
within the framework of passively evolving galaxies with AGB dust. Figure 16 shows the 
CFHT $r'$-band images of these galaxies. As the figure shows, their morphologies are 
predominately early-type. Therefore, these are passively evolving early-type galaxies.

The second class comprises optically red galaxies with $-$1 $<$ $N3-S11$ $<$ 0 (intermediate MIR-excess 
galaxies, hereafter intermediate-MXG). The SEDs suggest that they are either relatively young (compared to 
the weak-MXG), passively evolving galaxies ($\le$ 2-5 Gyr) or very weakly SF galaxies (around 
36\% of them have log(SSFR) $>$ -10.7, but are very near the cut in Fig. 15). If left alone, 
they are most likely to evolve into old, passively evolving galaxies. Figure 17 shows sample
images of these galaxies, which are mostly early types 
(i.e. ~71\% of the ANWS intermediate-MXGs are early types). 

The optically red galaxies with $N3-S11$ $>$ 0  (strong MIR-excess galaxies, hereafter strong-MXG) 
consist of weak SF galaxies with log(SSFR) $<$ -10 ($\sim$45\%, hereafter weak-SFG) and dusty SF 
galaxies with log(SSFR) $>$ -10  ($\sim$55\%, hereafter dusty-SFG). Interestingly, the mean value 
of log(SSFR) for dusty-SFG is comparable to that of blue SF galaxies (hereafter blue-SFG) with 
log(SSFR) of -9.7 versus -9.5, and higher than weak-SFG of -10.3. Both blue-SFG and dusty-SFG 
are thus vigorously SF galaxies. They may eventually stop star formation and evolve into the 
earlier-types. Figures 18 and 19 show the r$'$-band images of weak-SFG and dusty-SFG, respectively, 
which suggest that they have disk-like morphology. Interestingly, more than half of dusty-SFG are 
edge-on disks, indicating that half of those are viewed at higher disk inclination. Thus, we can
speculate that the dusty-SFG are strong SF late-type galaxies, but optically reddened.
On the other hand, weak-SFG have relatively lower SSFR than the blue-SFG, thus we can expect that their 
current/recent star formation is insufficient to change their optical color. In other words, in the
weak-SFG, the dominant stellar populations are generally old (no hot young stars, i.e. the average 
stellar age is greater than 1 Gyr) although their MIR-red colors ($N3-S11$ $>$ 0) indicate that they 
have recently formed some stars ($<$ 1 Gyr traced by P03 model) and/or
have a low level of ongoing star 
formation. Moreover, their dominant disk-like morphology (i.e. spirals) also indicates 
that a significant SFA could not have stopped very long ago, because spiral features fade within a few 
Gyrs without an interstellar gas supply (Lin \& Shu 1964; Bekki et al. 2002; Masters et al. 2010). 
It is therefore possible to conclude that, in terms of morphological transformation, weak-SFG are  
precursors of the the second class (intermediate-MXG). This is supported by the recent result of 
Hopkins et al. (2009) who suggest that disk galaxies with small gas fractions are more likely to be 
transformed into spheroids through minor mergers.  

Our classification scheme in 4 different populations based on the MIR color, 
is summarized in Table 4.
However, dusty-SFG could be an edge-on subsample of the blue SF galaxies because 
both have similar SSFR, indicating that both populations are at the same evolutionary stage. 
The only difference is that dusty-SFG are more likely to be edge-on. Therefore we divide 
galaxies into passively-evolving population (``weak-MXG''), transition populations 
(``intermediate-MXG'' and ``weak-SFG''), and strong star-forming populations 
(``dusty-SFG'' and ``blue-SFG'' ).


\begin{deluxetable}{cccccc}
\tabletypesize{\scriptsize} 
\tablewidth{0pc} 
\tablecaption{Galaxy classification.}

\tablehead{
  Galaxy type   &  Optical color   &  IR color     &  log(SSFR [yr$^{-1}$])  & Morphology fraction & Comments \\
(1) & (2) & (3) & (4) & (5)}

\startdata

weak-MXG          & red  & N3-S11 $<$ -1       & -11.2                    & early type ($>$ 90\%) & passively-evolving galaxies \\
intermediate-MXG  & red  & -1 $<$ N3-S11 $<$ 0 & -10.8                    & early type ($>$ 71\%) & transition populations \\
weak-SFG          & red  & N3-S11 $>$ 0        & -10.3 (-10.7$\sim$-10.0) & late type  ($>$ 67\%) & transition populations \\
dusty-SFG         & red  & N3-S11 $>$ 0        & -9.7 (-10.0$\sim$-9.0)   & late type  ($>$ 88\%) & SF galaxies \\
blue-SFG          & blue & N3-S11 $>$ 0        & -9.5 (-10.2$\sim$-8.9)   & late type  ($>$ 93\%) & SF galaxies \\

\enddata

\begin{flushleft}
Note. $-$ Col. (1): Classified galaxy type. 
Col. (2): First, the optical color-magnitude relation (CMR) is used to separate $red$ from $blue$ galaxies.
Col. (3): Second, the $red$ galaxies are subdivided by NIR$-$MIR ($N3-S11$) color. 
Col. (4): The mean values of specific star formation rate (SSFR) in units of dex. 
For SF populations, the range of SSFR is enclosed in brackets.
Col. (5): The dominant morphology from visual classification, and the fractions of
early types: 0, 1 and late types: 2, 3, 4 (see the caption in Fig. 16). \end{flushleft}
\normalsize

\end{deluxetable}


\begin{figure}
\epsscale{1}
\plotone{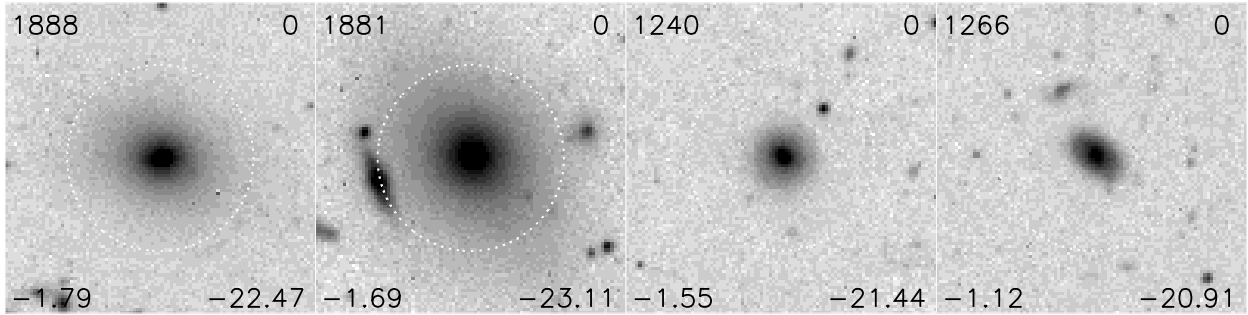}
\caption{Optical (CFHT $r'$ band) images (60$''$ $\times$ 60$''$) of the most massive galaxies with
         optically red and $N3-S11$ $<$ -1 (weak-MXG).
         Each image lists the object id, morphology, absolute N3 magnitude, 
         and $N3-S11$ color, on the top-left, top-right, bottom-right, and bottom-left, respectively.
         The morphology is classified by eye: 0 = bulge-dominated,  1 = bulge-dominated with disk feature,
         2 = disk-dominated with symmetric spiral, 3 = disk-dominated with asymmetric spiral,
         4 = others (merging system or unclassified). }
\end{figure}

\begin{figure}
\epsscale{1}
\plotone{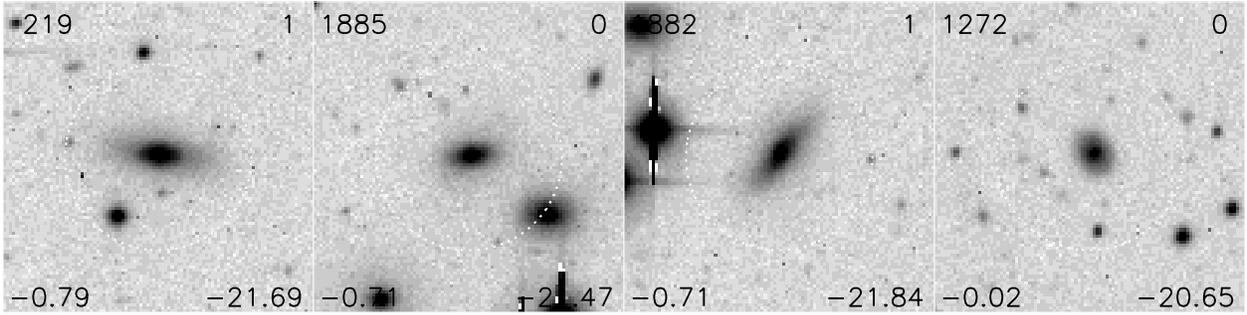}
\caption{Same as Fig. 16, but for the galaxies with
         optically red colors and 0 $<$ $N3-S11$ $<$ -1 (intermediate-MXG).}
\end{figure}

\begin{figure}
\epsscale{1}
\plotone{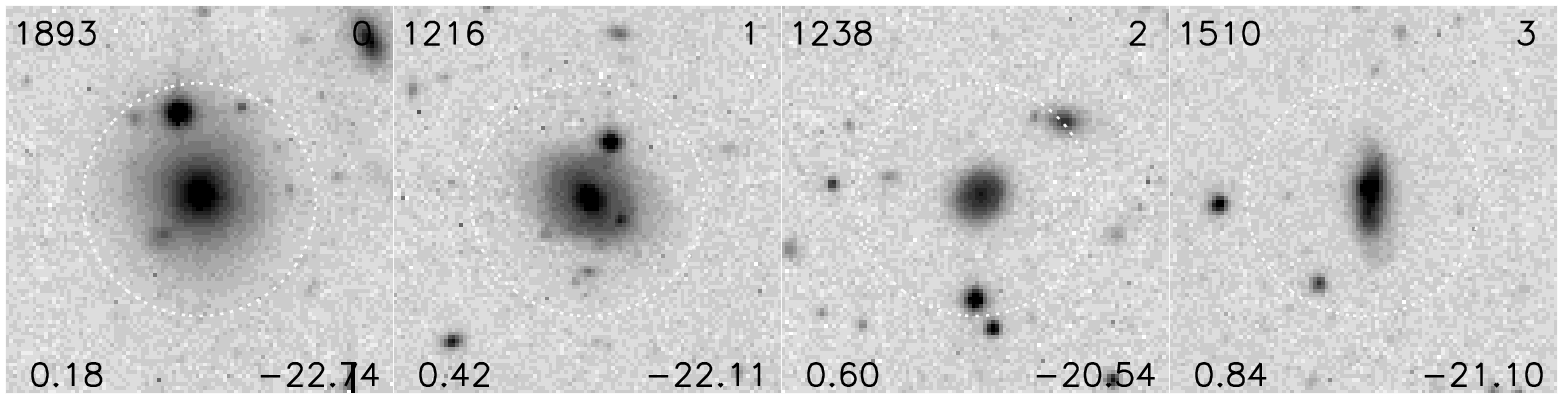}
\caption{Same as Fig. 16, but for the galaxies with 
         optically red colors and $N3-S11$ $>$ 0 (log SSFR $<$ -10; weak-SFG).}
\end{figure}

\begin{figure}
\epsscale{1}
\plotone{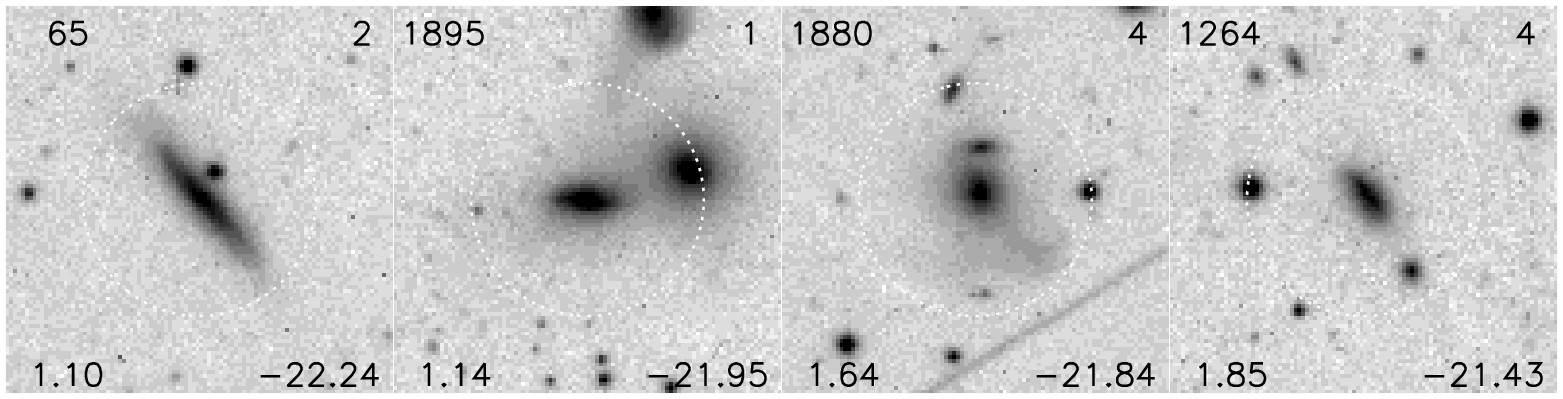}
\caption{Same as Fig. 16, but for the galaxies with 
         optically red colors and $N3-S11$ $>$ 0 (log SSFR $>$ -10; dusty-SFG).}
\end{figure}

\begin{figure}
\epsscale{1}
\plotone{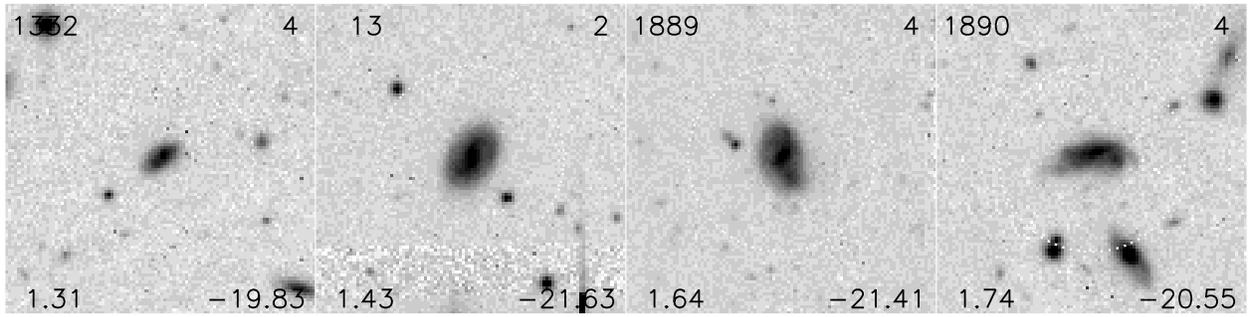}
\caption{Same as Fig. 16, but for the galaxies with 
         optically blue colors and $N3-S11$ $>$ 0 (blue-SFG).}
\end{figure}
\clearpage

\subsection{Transition populations}

\subsubsection{weak SF galaxies (weak-SFG)}

The weak-SFG in our sample show low levels of SFA (on average $\sim$ 4 times smaller SFRs 
than the blue-SFG at the fixed mass range log $M_{*}$/M$_{\odot}$ $=$ [9.5, 11]). However, 
their optical red colors are mostly due to their underlying old stellar populations, and
not an effect of dust reddening. In other words, they have older stellar populations and 
smaller SFRs (insufficient to change the optical color) 
compared to blue-SFG. Furthermore, they have mostly disk-dominated morphologies. Therefore, 
in terms of SFA in galaxy evolution, our weak-SFG are possible candidates for the 
transition population in which star formation is suppressed, compared to blue-SFG, and 
their red colors are due to old stellar populations. 
These populations are very similar to `anemic spirals' discovered by 
van den Bergh (1976) which are thought to be in the transition from blue, SF field spirals to
red, non-SF cluster S0 galaxies.
They are also generally at a similar evolutionary stage to the transition populations 
(red spirals or red SFs) selected using different criteria in other works 
using SDSS galaxies (e.g., Bamford et al. 2009; Masters et al. 2010) and galaxies in 
the Abell 901/902 supercluster at redshift $\sim$ 0.17 (e.g., Gallazzi et al. 2009; Wolf et al. 2009).

\subsubsection{intermediate MIR-excess galaxies (intermediate-MXG)}

The intermediate-MXG in our sample are defined as bluer in $N3-S11$ colors (indicating 
lower SSFRs) than weak-SFG, but redder (indicating younger ages) than weak-MXG. This implies 
that their MIR emission arises from very weak SFA (on average $\sim$ 3.2 times lower SFR 
than weak-SFG), or AGB stars with intermediate ages of 1 $\sim$ 5 Gyr (i.e. past SFA). However, 
their optical red colors are definitely from an old stellar population, because their 
estimated mean value of E$_{B-V}$ ($\sim$ 0.05) is similar to weak-MXG.
Although the origin of the MIR emission is complex, the mean stellar 
ages of intermediate-MXG traced by the amount of warm dust emission per stellar mass are obviously 
older than the ages of the weak-SFG and younger than the weak-MXG. Thus, it is possible that this 
population is on the evolutionary stage of migration to weak-MXG from weak-SFG, or directly from 
blue-SFG. Furthermore, in contrast to weak-SFG, the dominant morphologies of intermediate-MXG are 
early-types (more than 71\%), 
supporting the idea that these are in the final transition phase just before passively becoming 
weak-MXG. Therefore, our intermediate-MXG represent populations where major SFA ended over 1 Gyr ago.
The MIR-weighted mean stellar ages for these galaxies are 1-5 Gyrs, if we assume single-burst stellar 
populations. This result can be explained if the MIR-excess emission of red, early-type galaxies is 
mainly from AGB stars (Ko et al. 2009). It is supported by Vega et al. (2010), who suggest 
that the unusual PAH emission in the 6$-$14 $\mu$m region in nearby early-type galaxies arises from 
intermediate-age carbon stars, and they can be formed by rejuvenation episodes within the last few Gyr at 
the 1\% level of the mass of the galaxy. 

Our intermediate-MXG seem to be closely related to the UV-excess galaxies in Yi et al. (2005). They found 
that roughly 15\% of nearby early-type galaxies have excess NUV emission indicating recent ($<$ 1 Gyr) 
star formation comprising 1\%$-$2\% of the total stellar mass. 
Around 1 Gyr after a single-burst 
of star formation, massive stars (O, B, and A stars) have expired, and recent star formation 
indicators (e.g., NUV flux and H$\beta$ line index) are no longer good
tracers of the star formation 
history. However, the MIR-excess emission over stellar light can trace star formation over a much longer 
period, since low to intermediate mass (1 $-$ 9 M$_{\odot}$) stars evolve to the AGB phase, and their 
circumstellar dust emission is strong in the MIR. Frogel et al. (1990) found that the contribution of 
AGB stars to the bolometric luminosity peaks at more than 40\% at ages from 1.1 to 3.3 Gyr, but 
rapidly falls to less than 5\% at 10 Gyr. In other words, our intermediate-MXG seem to be a population
of descendants of objects showing recent SFA within $\sim$1 Gyr, which are UV-excess early-type 
galaxies and post-starburst (E+A) galaxies (Choi et al. 2009). However, we expect that those young 
early-type galaxies have different evolutionary histories from our
weak-SFG, due to their different morphologies.

\section{ENVIRONMENTAL AND MASS DEPENDENCE OF GALAXIES IN TRANSITION PHASE}

We now investigate how the properties of galaxies are influenced by their stellar mass and local 
environment. We focus on the intermediate-MXG and the weak-SFG as objects that 
might be in the evolutionary transition from blue, star-forming late-types to red, passive early-types.

If the properties of galaxies are affected by environmental mechanisms, such as hydrodynamic or
gravitational processes (e.g., Boselli \& Gavazzi 2006 and Park \& Hwang 2009 for a review), we expect 
to see an increase of transition populations in specific environments. For example, when blue spiral 
galaxies infall from low- to high-density regions, (although it depends on the relevant time-scale) 
both global halo properties and/or the local galaxy density can produce a gradual/sharp decline of 
star formation and morphological transformation by external factors, such as tidal interactions and 
gas stripping. As a result, SF galaxies in dense regions show lower SFR, and the fraction of early-type 
galaxies increases with increasing density or toward the center of the galaxy cluster. 

On the other hand, if the effects of stellar mass on the galaxy properties are stronger than those of
environment, then we would not expect any significant changes in the distribution of transition galaxies 
over a range of local density at fixed stellar masses.
In this case, the stellar mass is the primary parameter governing the suppression of star formation 
and the transformation of morphology, and thus transition populations would only show a trend with their 
stellar mass.

In an attempt to analyze the dependence of the properties on the stellar mass and the environment 
independently, we plot the change in the relative fraction of different types of galaxies as a function 
of the stellar mass at low, intermediate, and high density regions (Fig. 21), and as a function of the 
local density at different mass bins (Fig. 22). Here, $\Sigma$ is used for the local density, as 
described in Section 3.2. 
The type fraction ($f_{t}$) is defined as the ratio between the number
$N_{t}$ of each galaxy-type to the 
$N_{total}$ total number galaxies in fixed mass and density bins. We include $S11$-undetected samples to 
calculate $N_{total}$. The uncertainties of the fractions are estimated by calculating the variance 
in the likelihood of the fraction (e.g., De Propris et al. 2004). Assuming that the fraction has 
the form of the likelihood function 

$L \propto f_{t}^{N_{t}}(1-f_{t})^{N_{total}}$ \\ 
and its maximum is $\frac{N_{t}}{N_{total}}$, the variance of the fraction is 

$\sigma^{2}(f_{t}) = (\frac{d^{2}lnL}{df_{t}^{2}})^{-1}$ \\ \\
when the likelihood function has a Gaussian form. Note that our estimate of the standard deviation
might be low due to the distribution of our small samples (i.e. the likelihood function may not be
Gaussian). 
In order to avoid the lower detection rate of galaxies with $N3-S11$ $<$ 0 
(only the weak-MXG and intermediate-MXG are contained in our $N3$-selected sample) due to the $S11$ 
detection limit, Figure 23 shows galaxies at stellar mass bin log $M_{*}$/M$_{\odot}$ $=$ [10, 11]. 
We include $S11$-undetected samples as part of the weak-MXG since their SSFRs are low enough.
Table 5 summarizes the results in Figure 23. From Figs. 21$-$23 and table 5, 
our results can be summarized as follows.

\begin{figure}[ht!]
\epsscale{1}
\plotone{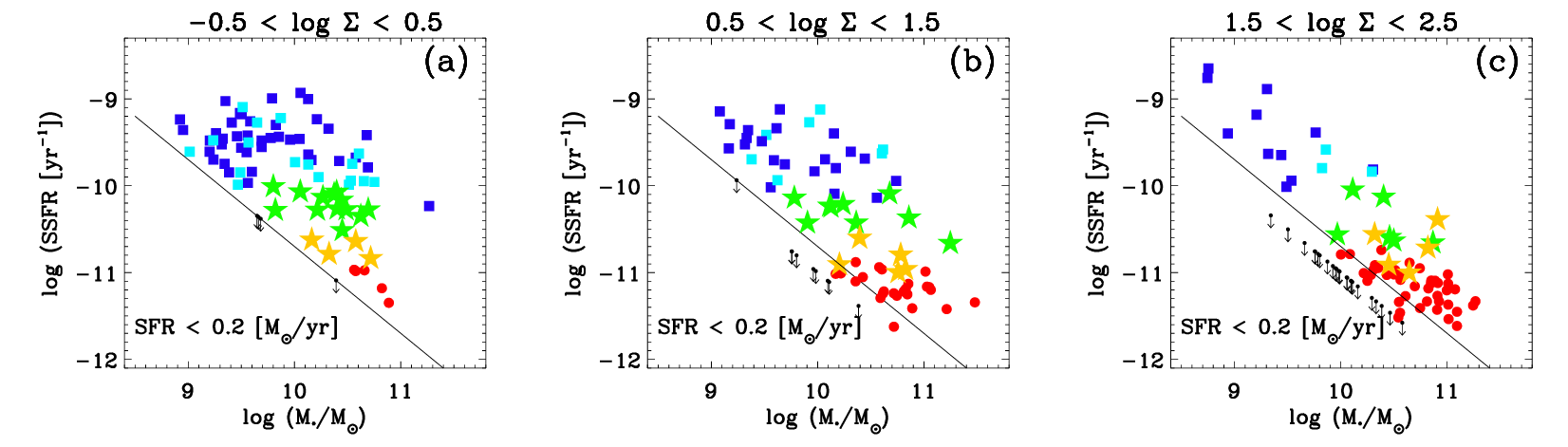}
\plotone{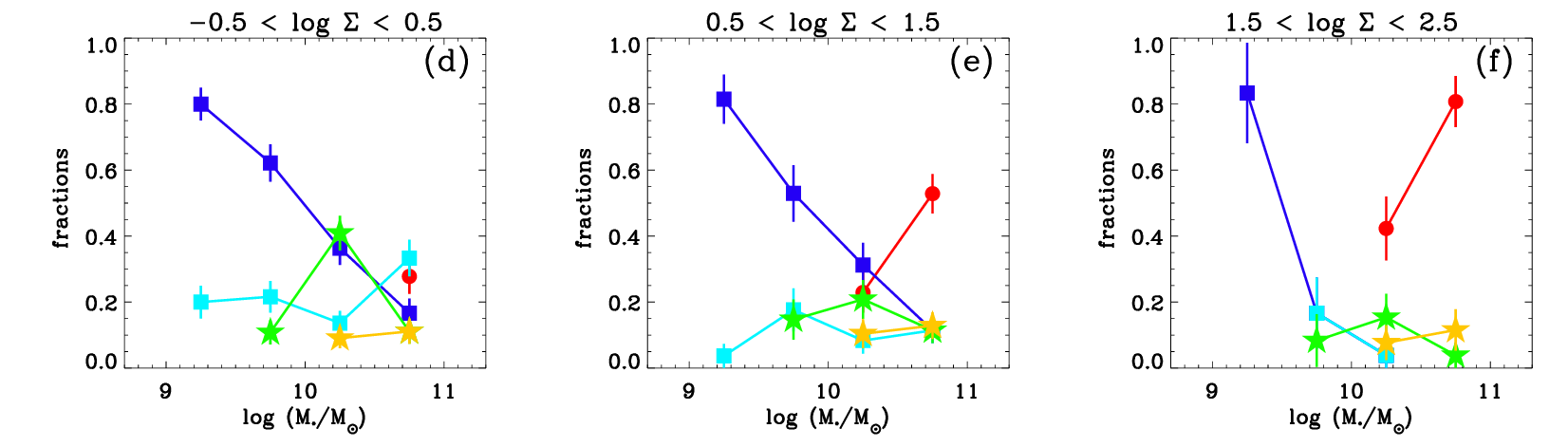}
\caption{\textit{Upper}: SSFR as a function of stellar mass in three density bins for our $N3$-selected
                         supercluster member galaxies in A2255 and the ANWS. Local density is increasing 
                         from left to right. The red circle, yellow star, green star, cyan square, and 
                         blue square represent weak-MXG, intermediate-MXG, weak-SFG, dusty-SFG, and 
                         blue-SFG, respectively. Circles, stars, and squares also represent 
                         passively-evolving, transition, and strong SF populations, respectively.
                         The small circle with an arrow represents the
                         object that is undetected in $S11$ 
                         (see Fig. 13), and the solid lines in each panel show the detection limit 
                         of 0.2 M$_{\odot}yr^{-1}$ for the ANWS sample.
         \textit{Lower}: The fraction of each galaxy-type as a function of stellar mass.
                         Type fractions are defined as the ratio
                         between the number galaxies in each type
                         class and the total number of galaxies at a
                         fixed mass. The errors are calculated
                         assuming that the fractions are the maximum values of the likelihood function 
                         (see text for details). The total number
                         galaxies includes objects that are undetected
                         in $S11$. Mass bins with less than three galaxies are excluded.}
\end{figure}

\begin{figure}[ht!]
\epsscale{1}
\plotone{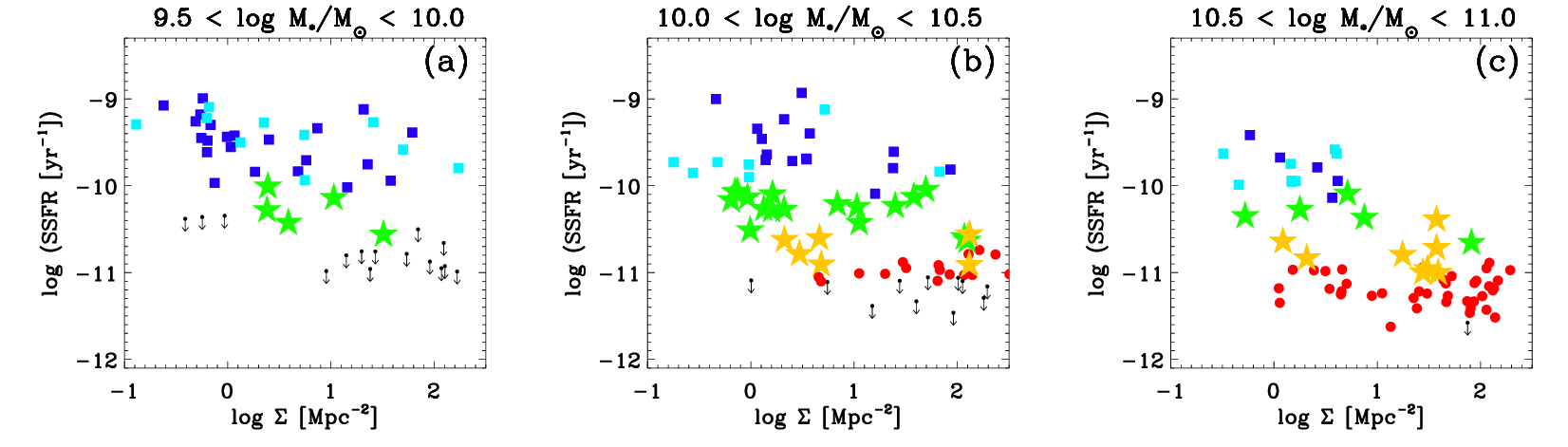}
\plotone{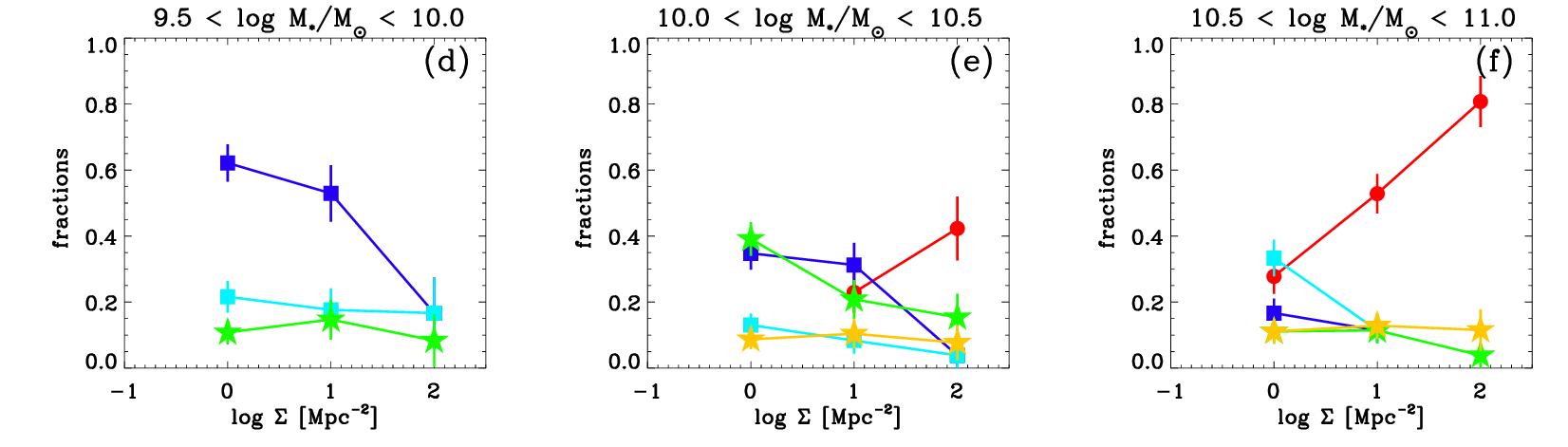}
\caption{\textit{Upper}: SSFR as a function of galaxy local density in three stellar mass bins
                         for our $N3$-selected supercluster member galaxies in A2255 and the ANWS.
                         The symbols are the same as in Fig. 21.
         \textit{Lower}: The fraction of each galaxy type as a function of local density.}
\end{figure}

\begin{itemize}
\item   We find that the weak-SFGs are mostly dominant at mass bin log $M_{*}$/M$_{\odot}$ $=$ [10, 10.5] 
        and at density bin log $\Sigma$ = [-0.5, 0.5] (Fig. 21(d) and Fig. 22(e)). In this mass 
        range (Fig. 22(e)), more than 20\% of the population in
        all density bins are the weak-SFG, and, in particular,
        the weak-SFG are the dominant population in the density bin of 
        log $\Sigma$ = [-0.5, 0.5] ($\sim$ 40\%). Interestingly, this density bin corresponds to the 
        region in the outskirts of A2255 and the intermediate-density regions of the ANWS. 
        This is confirmed in Figure 24, which shows the distribution
        of each galaxy-type only for the most
        massive galaxies (log $M_{*}$/M$_{\odot}$ $>$ 10) on the
        number density maps of all member galaxies. 
        The weak-SFG tend to avoid the central region of A2255 (the
        two weak-SFG close to the center have redshifts of
        0.07379 and 0.07387, placing them at the edge of the
        velocity distribution of A2255, so that their location is
        likely a projection effect).
        However, when galaxies in both more-massive (log $M_{*}$/M$_{\odot}$ $=$ [10.5, 11.0]; Fig. 22(f)) 
        and less-massive (log $M_{*}$/M$_{\odot}$ $=$ [9.5, 10.0]; Fig 22(d)) bins are
        considered, the weak-SFG are not as significant,
        contributing with less than 20\% of the population in any
        density bins, and their environmental trend is weaker. 
        Note that, although the weak- and intermediate-MXG are not
        detected in the lowest mass bin because of
        the $S11$ detection limit, our result matches well the trend of red spirals of 
        Wolf et al. (2009; see their Fig. 15).         

\begin{figure}[ht!]
\epsscale{1}
\plotone{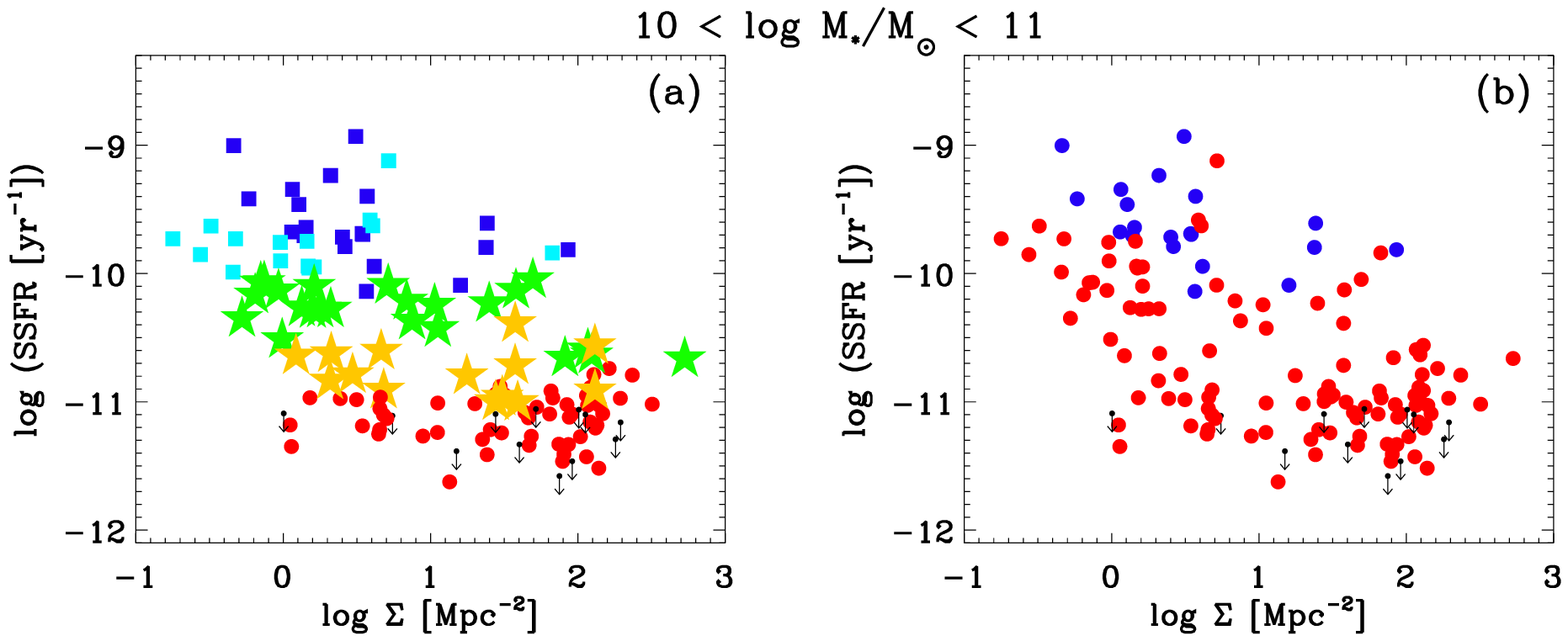}
\plotone{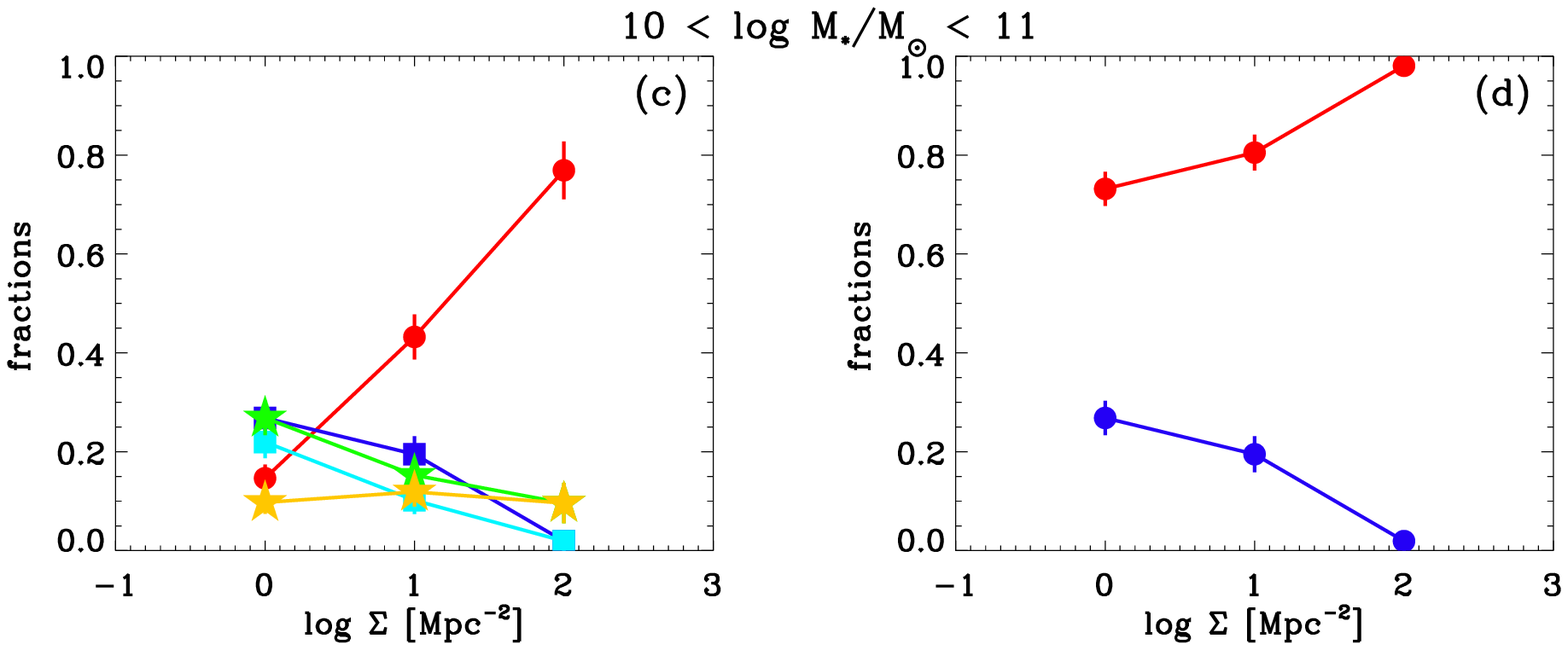} 
\caption{\textit{Upper}: SSFR as a function of galaxy local density at stellar mass bin 
                         log $M_{*}$/M$_{\odot}$ $=$ [10, 11] for our $N3$-selected supercluster member 
                         galaxies in A2255 and the ANWS. In the left panel, each galaxy-type 
                         is the same as in Fig. 21, while red and blue circles in the right panel 
                         indicate optical red and blue galaxies which are divided using the 
                         $B-R$ vs. $M_{N3}$ CMR (see Fig. 13 for details).
         \textit{Lower}: The fraction of each galaxy type as a function of local density.
                         Undetected samples in $S11$ are counted as the weak-MXG ($left$) and red ($right$)
                         due to their low SSFR.}
\end{figure}

\begin{deluxetable}{cccc}
\tabletypesize{\scriptsize} 
\tablewidth{0pc} 
\tablecaption{The fraction of each galaxy-type in Figure 23(c). \label{tab-numden}}

\tablehead{   &    \multicolumn{3}{c}{log $M_{*}$/M$_{\odot}$ $=$ [10, 11]}  \\
              \cline{2-4} \\
  Galaxy type   &  low-density (\%)                &  intermediate-density (\%)       &  high-density (\%)   \\
                &  (-0.5 $<$ log $\Sigma$ $<$ 0.5) &  (0.5 $<$ log $\Sigma$ $<$ 1.5)  &  (1.5 $<$ log $\Sigma$ $<$ 2.5)}
\startdata

weak-MXG                & 14 & 43 & 76  \\
intermediate-MXG        & 10 & 12 & 10  \\
weak-SFG                & 27 & 15 & 10  \\ 
dusty-SFG               & 22 & 10 & 2    \\
blue-SFG                & 27 & 20 & 2   \\

\enddata

\end{deluxetable}

\item   At log $M_{*}$/M$_{\odot}$ $>$ 10, the overall fraction of SF galaxies 
        (blue-, dusty-, and weak-SFG) decreases gradually by $\sim$ 76\%, $\sim$ 45\%, and 
        $\sim$ 14\% in low-, intermediate-, and high-density bins respectively (see Fig. 23(c)). 
        The overall fraction of red galaxies increases increasing local 
        density (see Fig. 23(d)). This is basically a confirmation of the SFR--density relation 
        (e.g., Lewis et al. 2002; Kauffmann et al. 2004; Weinmann et al. 2006; Hwang et al. 2010) 
        and CDR (e.g., Pimbblet et al. 2002; Blanton et al. 2005; 
        Cucciati et al. 2006). This trend indicates that massive galaxies are mostly red in all 
        environments, and comprise most of the galaxies in the highest density regions.
        However, if optically red galaxies 
        are divided into four different sub-populations, then the interpretation needs 
        to be done carefully. If we focus on the weak-MXG (Fig. 23(c)), there is a strong environmental 
        dependence, there being a very small fraction of weak-MXGs in the lowest density bin. 
        However, the fraction of red galaxies presents a high value ($\sim$ 73\%, 
        see Fig. 23(d)) in the lowest density bin due to the rather
        high proportion of SF galaxies (weak- and dusty-SFG) that are red.

\begin{figure}[ht!]
\epsscale{1}
\plotone{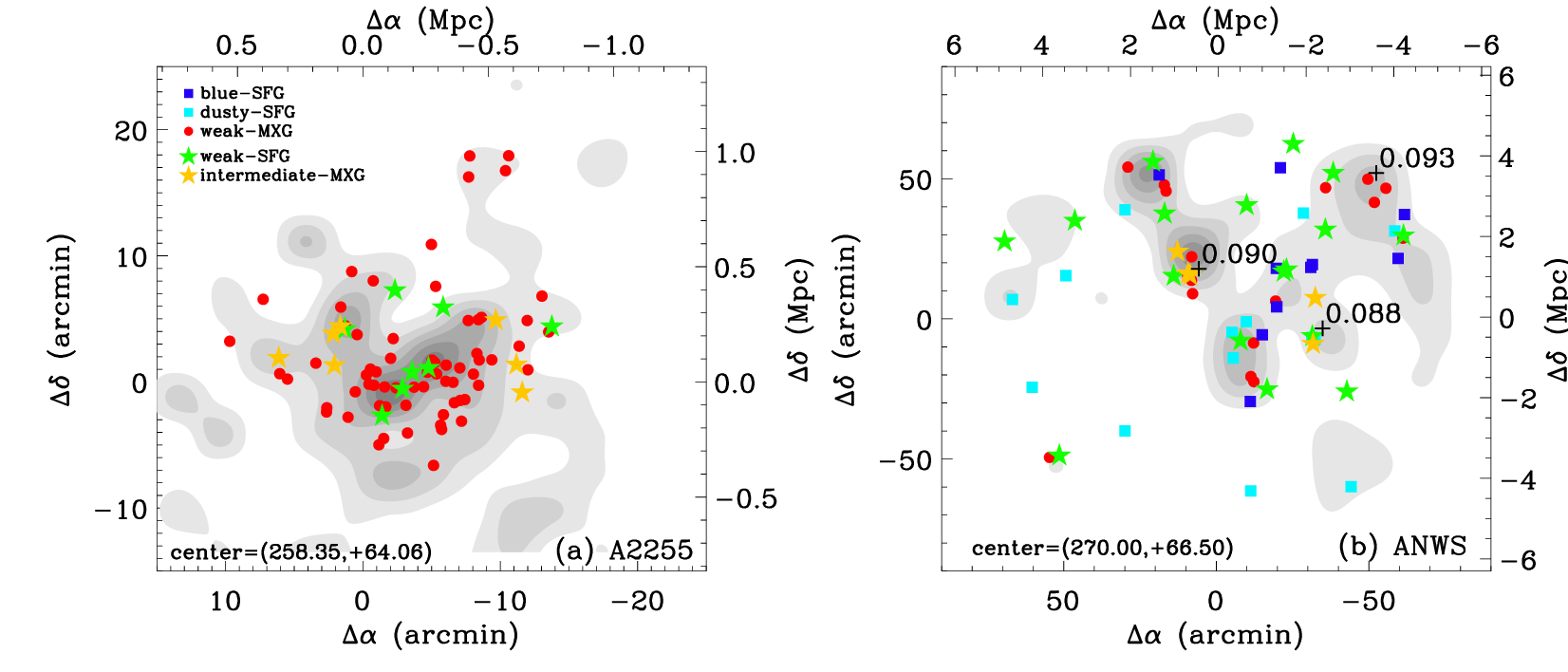}
\caption{Spatial distribution of each galaxy type in Fig. 23, on the smoothed galaxy number density maps 
         for all supercluster member galaxies in A2255 ($left$) and in the ANWS ($right$). 
         In the right panel, the three ``+'' signs indicate the center of X-ray detected groups with
         the mean redshifts given (Henry et al. 2006). North is up, and east is to the left.}
\end{figure}
        
\begin{figure}[ht!]
\epsscale{1}
\plottwo{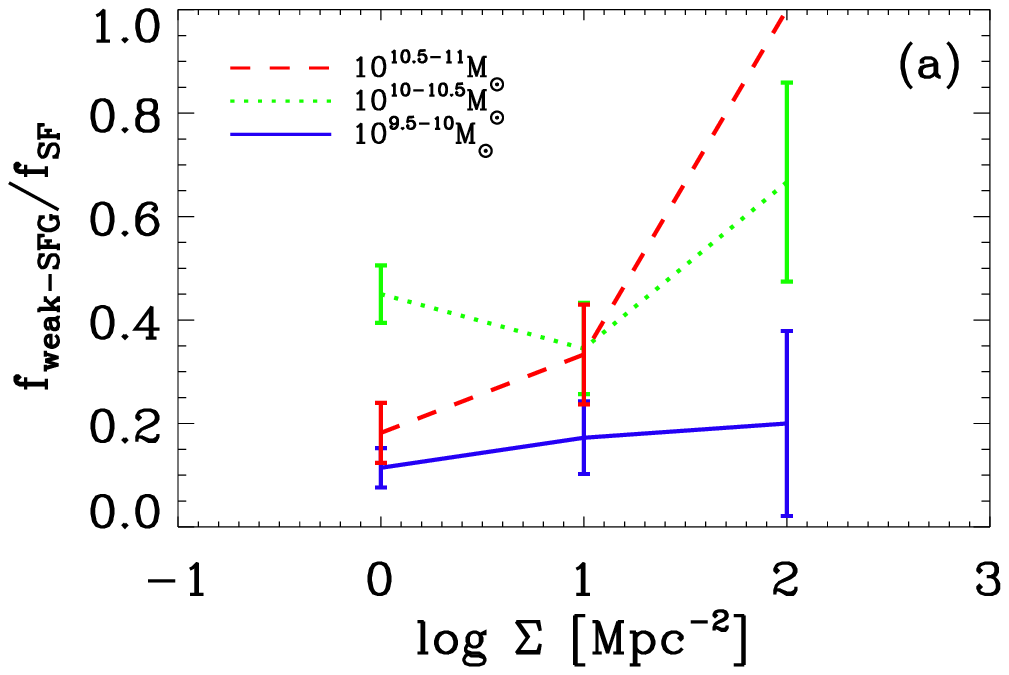}{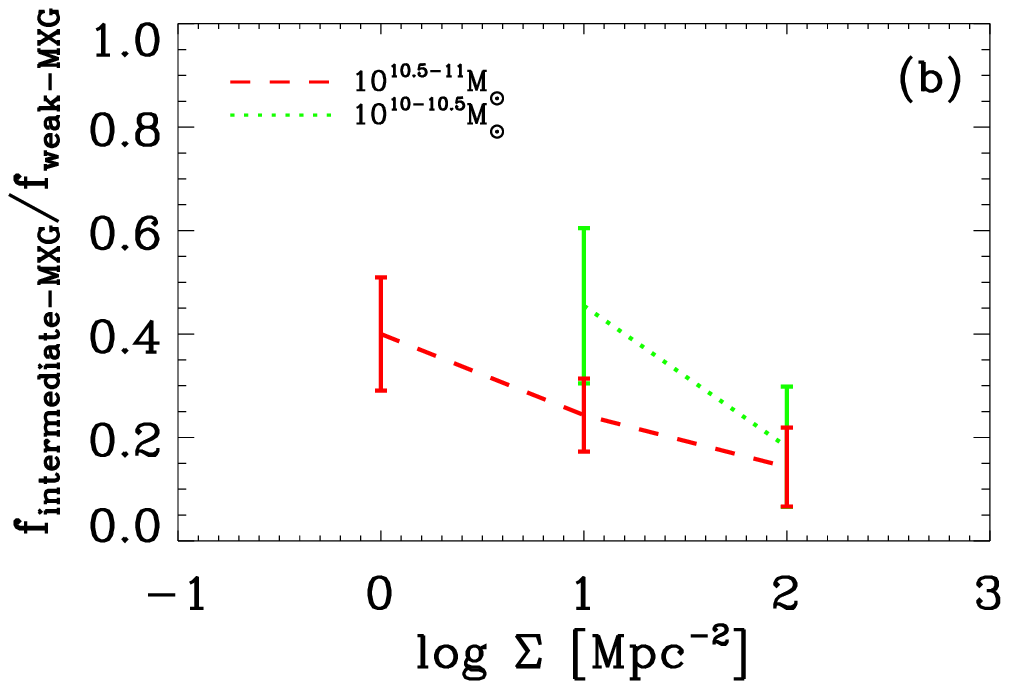}
\caption{\textit{Left}: The fraction of weak-SFG among SF galaxies (weak-, dusty-, and blue-SFG) 
                        is plotted as a function of local density in three stellar mass bins: 
                        log $M_{*}$/M$_{\odot}$ $=$ [9.5, 10], [10, 10.5], [10.5, 11].
         \textit{Right}: The relative fraction of the intermediate-MXG versus the weak-MXG
                         is plotted as a function of local density in two stellar mass bins: 
                         log $M_{*}$/M$_{\odot}$ $=$ [10, 10.5],
                         [10.5, 11]. Because of the cut in 
                         $S11$, less-massive (log $M_{*}$/M$_{\odot}$ $<$ 10) MIR-blue ($N3-S11$ $<$ 0) 
                         galaxies are not detected.}
\end{figure}

\item   The relative fraction of weak-SFGs among SF galaxies (weak-, dusty-, and blue-SFG),
        as a function of local density at fixed mass, allows us to study the importance of 
        mass/environment in SF quenching. In Figure 25(a), at all local densities, the relative 
        weak-SFG fraction is higher at large stellar mass (log $M_{*}$/M$_{\odot}$ $>$ 10) than at 
        low stellar mass (log $M_{*}$/M$_{\odot}$ $<$ 10). 
        The fraction of weak-SFGs is $\sim$ twice as high 
        at the highest density bin, compared with the lowest density
        bin when only massive galaxies are considered. 
        However, for less-massive galaxies, the fraction does not change significantly with the local 
        density. This suggests that the weak-SFG are likely to be more massive than strong SF 
        galaxies (dusty- and blue-SFG) and comprise a significant fraction of all massive SF galaxies in 
        high-density environments.

\item   We find roughly 10\% intermediate-MXG among all massive 
        (log $M_{*}$/M$_{\odot}$ $>$ 10; Fig. 23(c)) galaxies in all density bins. However, the relative 
        fraction of these versus the weak-MXG decreases as the density increases, by $\sim$ 42\%, 
        $\sim$ 22\%, and $\sim$ 12\%, respectively. Thus about half of
        the massive early-type galaxies on the red-sequence at the outskirts of the cluster show significant 
        excess MIR emission compared to normal (passively evolving) early-type galaxies. 
        In Figure 25(b), when the intermediate-MXG are divided into two mass bins 
        (log $M_{*}$/M$_{\odot}$ $=$ [10, 10.5], [10.5, 11]), the relative fraction of the 
        intermediate-MXG is higher in the lower-mass bin. It is thus possible to infer that 
        the intermediate-MXG are less-massive and tend to be located in the outer parts of the cluster. 
        This is largely consistent with our study of the galaxies in A2218 (Ko et al. 2009). 
        Although the size of our mass-limited sample of intermediate-MXG is too small to determine 
        the mass-dependence for this population,
        we can see the environmental-dependence for massive galaxies.
        In Figure 24, the intermediate-MXG are likely to be located in the outskirts of the 
        cluster and near group centers (corresponding to the lowest density bin in Fig. 25(b)). 
        Therefore, an environmental action is necessarily required to explain the properties of 
        this population. It is expected that the intermediate-MXG evolve into weak-MXG.
        Furthermore, SSFRs (approximately MIR-weighted mean stellar ages) of the intermediate-MXG 
        are much smaller than those of weak-SFG, indicating that the weak-SFG are at an earlier 
        evolutionary stage than the intermediate-MXG. Also, the main difference between both transition 
        populations is the morphology (disk-dominated for weak-SFG and bulge-dominated for intermediate-MXG).  

\end{itemize}

\section{DISCUSSION}

Recent studies of red galaxies defined by optical CMR cuts indicate
that they contain several populations at different evolutionary stages (e.g., Lee et al. 2008; 
Ko et al. 2009; Cortese \& Hughes 2009; Wolf et al. 2009; Gallazzi et al. 2009; 
Tran et al. 2009; Bamford et al. 2009; Bundy et al. 2010; Masters et al. 2010; 
Salim \& Rich 2010). For example, red, early-type galaxies are found to have a 
wide range of MIR-excess of non-stellar origin, suggesting that some of these 
experienced recent star formation episodes. In addition, some red-sequence galaxies are found 
to have UV excess suggesting weak SFA.

We expect that there are two different phases (star formation quenching and morphology 
changing), when a blue, star-forming, late-type galaxy turns into a red, quiescent, 
early-type galaxy. Recent studies revealed that star formation quenching (optical 
color change) is not always accompanied by morphological change (e.g., Blanton 
et al. 2005, S\'{a}nchez et al. 2007, Bamford et al. 2009; Wolf et al. 2009). 
In other words, the time scale of transition from blue to red and of morphological 
change from late-type to early-type is different, and seems to be a function of stellar 
mass and local galaxy density. Observationally, the existence of red spirals (i.e. weak-SFG) 
and blue early-type galaxies, and their preference for specific masses and local densities 
supports this idea. 

To trace mass- and environment-dependence of changes in color and morphology, we focused on 
two different categories of transition galaxies (intermediate-MXG and weak-SFG).  Specifically, their 
NIR-MIR color ($N3-S11$)  is a good tracer of SSFR, and the SSFR of SF galaxies is 
not sensitive to their mass and environment (e.g., Peng et al. 2010).

\subsection{Transition populations I: weak-SFG} 

Our result that the weak-SFG are mainly dominant at intermediate mass 
(log $M_{*}$/M$_{\odot}$ $=$ [10, 10.5]) in the cluster outskirts agrees with previous studies: 
both Wolf et al. (2009) and Masters et al. (2010) show that red spirals are predominant at 
intermediate local density (infall regions of clusters), and at the
higher mass end ($>$ 10$^{10}$ M$_{\odot}$).
Quantitatively, we find that the fraction of weak-SFG among SF galaxies (blue-, dusty-, and weak-SFG) range 
from 71\% at the cluster core to 36\% in the outskirts of the cluster, at log $M_{*}$/M$_{\odot}$ $=$ [10, 11]. 
This suggests that the suppression of star formation progresses rapidly in high-density environments, so 
that the weak-SFG shifted into quiescence earlier than those in the
lower density environments, at 
a fixed mass ($>$ 10$^{10}$ M$_{\odot}$). This suggests the acceleration of ``downsizing'' in overdense regions 
(e.g., Bundy et al. 2006). If we assume the ``downsizing'' scenario, then galaxies with the same mass and 
SSFR should have similar star formation histories. However, the much higher weak-SFG fraction among SF galaxies 
and much lower weak-SFG fraction among red galaxies in high-density
environments compared to low-density environments suggests that the star formation of their progenitor galaxies
(blue-SFG) could be suppressed efficiently because of the high galaxy number density and/or much
longer interaction with the cluster environment.
Thus massive weak-SFG in high-density regions could be already replaced by intermediate-MXG and/or weak-MXG.

In contrast, at the highest masses (log $M_{*}$/M$_{\odot}$ $=$ [10.5, 11]), the fraction of weak-SFG 
is $>$2 times smaller than the mass cut (log $M_{*}$/M$_{\odot}$ $<$ 10.5) in all density bins 
($<$ 10\% of all galaxies). This can be interpreted as a mass-dependent star formation history where 
massive galaxies are much older, became passive earlier than less massive galaxies, 
and have been undergoing a much higher frequency of mergers, so that their morphologies have
already transformed to early types.

At lower masses (log $M_{*}$/M$_{\odot}$ $=$ [9.5, 10]), even considering our $S11$ detection limit, 
the fraction of weak-SFG among SF galaxies appears to decrease sharply at all density bins, 
compared to larger masses (see the left panel of Fig. 25). This is consistent with the results of 
Wolf et al. (2009) where they suggest that the star formation of low-mass galaxies in clusters is 
suppressed quickly and the morphological change happens simultaneously, hence red spirals are very 
rare (see also Boselli et al. 2008).
Therefore, we confirm the previous results that low-mass blue-SFG infalling 
from outskirts of clusters have experienced halo and disk gas stripping via some environmental 
effects, and almost simultaneously spiral structures have disrupted into early-type morphology, in 
contrast to higher-mass galaxies whose spiral structures persist much longer. 
It is supposed that low-mass galaxies with depleted gas disks are susceptible to morphological change 
through minor merger events (Hopkins et al. 2009). Furthermore, Masters et al. (2010) also found that the 
fraction of red spirals with smaller masses ($<$ 10$^{10}$ M$_{\odot}$) is very low, in contrast to the 
larger masses.

From the behavior of the weak-SFG, star formation quenching is
affected both by the stellar mass
and environment. However, the environmental dependence works
differently in each mass bin.

\subsection{Transition populations II: intermediate-MXG} 

Another proposed transition population is the intermediate-MXG, which has been already classified 
in our analysis of A2218 and A2255 (Ko et al. 2009; Shim et al. 2011). These galaxies are optically 
red and have early-type morphologies, but show broad emission in the MIR 
(e.g., Bressan 2006). These suggest that the MIR-weighted mean stellar ages of these galaxies 
are younger than those of the weak-MXG. They also show a wide range of MIR-excess emission, suggesting 
a variety of star formation histories among red, early-type galaxies.
 
In the right panel of Figure 25, although we can only explore massive intermediate-MXG due to our $S11$ 
detection limit, we find that these galaxies are relatively low-mass systems among massive 
($>$ 10$^{10}$ M$_{\odot}$) red, early-type galaxies, and are likely to be located in the outer parts 
of the cluster. This is consistent with the results in A2218.

At larger masses ($>$ 10$^{10}$ M$_{\odot}$), about 51\% of the galaxies show quenched 
or decreased star formation (i.e. weak- and intermediate-MXG, and weak-SFG) and around 24\% 
have early-type morphology (i.e. weak-and intermediate-MXG) in the
outskirts of clusters;
when the cluster core is considered, these proportions increase to 96\% and 86\% respectively. 
Furthermore, with increasing local density, the relative
fraction of these transition galaxies versus the weak-MXG decreases
sharply, while versus the weak-SFG it increases. These trends can be explained if the morphological 
transformation starts at the outskirts and the process is mostly completed at high density, and 
the intermediate-MXG are a set of products of morphological transition between the weak-SFG and 
the weak-MXG at all cluster environments. From the behavior of the intermediate-MXG, morphological 
transformation of galaxies can be explained by the environmental effects. 

The behavior of our transition populations (weak-SFG and intermediate-MXG) suggests a 
possible scenario of evolutionary history of galaxies, from star-forming, late-type galaxies to 
non-star-forming, early-type galaxies. On this issue, based on the
findings above, we can speculate that 
as the gas supply decreases and the SFRs continue to decline, blue-SFG may naturally change to weak-SFG 
with gas-poor disks, mainly governed by their mass (massive galaxies evolve faster than less-massive 
ones in optical color change), and that this transition is accelerated in the high-density environment. 
Then a large fraction of massive weak-SFG proceed slowly through several environmental processes, particularly 
starting at the outskirts of clusters, while for less-massive galaxies this happens faster. Finally they 
transform into the weak-MXG through the intermediate-MXG. S\'{a}nchez et al. (2007) suggested a two-step 
scenario in which star formation is quenched first, and morphological transformation follows on longer 
timescale, from the analysis of A2218. S\'{a}nchez-Bl\'{a}zquez et al. (2009) also suggested that the 
timescale of morphological transformation of the galaxies entering the red-sequence is different from 
that of star formation quenching.

\section{SUMMARY AND CONCLUSIONS}
                              
We have investigated the MIR properties of optical red-sequence galaxies within 
a supercluster in the NEP region at redshift $\sim$0.087, using the ANWS (5.4 deg$^{2}$) 
and AKARI IR (2$-$24 $\mu$m) observations of A2255, in conjunction with NUV-optical SEDs 
and optical spectroscopy. AKARI 11$\mu$m flux traces not only 
the amount of recent SFA, but also the presence of intermediate age stellar populations 
(i.e. past SFA). Therefore, the NIR$-$MIR ($N3-S11$) color can be a good indicator of SSFRs, 
whereby we can identify dusty SF galaxies (dusty-SFG) and transition galaxies among 
red-sequence galaxies.

We find that $\sim$ 22\% of the massive ($>$ 10$^{10}$ M$_{\odot}$) red-sequence galaxies are 
dusty-SFG in low-density regions. Their SSFRs are comparable to those of blue-SFG and 
their environmental trend is similar to SF populations (blue- and weak-SFG). 
Thus their optical red colors are due to dust extinction or/and higher disk inclination.
  
We also find that red-sequence galaxies, excepting dusty-SFG, consist not only of passively
evolving galaxies, but also of weak-SFG (disk-dominated SF galaxies which have SSFR 
lower than blue-SFG), and intermediate-MXG (bulge-dominated galaxies showing broad 
non-stellar MIR emission compared to weak-MXG). These two populations may represent 
transition galaxies from blue, star-forming, late-type galaxies evolving into red, quiescent, 
early-type ones. In this study, we have focused on properties of these transition galaxies, and on how 
the fraction of transition galaxies depends on the stellar mass and the local density. 
Our main conclusions are summarized as follows.

\begin{itemize}
 \item The weak-SFG are found to be similar to the red spirals of the Galaxy Zoo (Masters et al. 2010) 
       and optically passive spirals (Wolf et al. 2009) in the A901/2 cluster. Consistent 
       with previous studies, at the same mass range (log $M_{*}$/M$_{\odot}$ $=$ [10, 11]), 
       they show a lower level of SFA (on average $\sim$ 4 times lower) than blue-SFG.
\end{itemize}
       
\begin{itemize}
 \item The intermediate-MXG show little recent ($<$ 1 Gyr) SFA enough to 
       contribute to the total NUV flux, and their MIR luminosity-weighted mean stellar ages (1--5 Gyr) 
       are older than the weak-SFG and younger than the weak-MXG.
\end{itemize}

\begin{itemize}
 \item In the evolution of a galaxy, the weak-SFG could be candidates for the transition 
       stage between blue-SFG and intermediate-MXG, where the star formation is quenched,
       while the intermediate-MXG are likely to be placed in an intermediate stage between 
       blue/weak-SFG and weak-MXG, where the morphology is transformed into early types.
\end{itemize}

\begin{itemize}
 \item The transition population is the most abundant at intermediate local densities 
       (outskirts of clusters), suggesting that most of the
       action takes place at intermediate densities.
       The relative fraction of weak-SFG versus intermediate-MXG increases as
       local density decreases. This indicates that the star formation quenching is 
       ongoing at the outskirts, and the process is mostly completed at high density. 
       The morphologies of the intermediate-MXG are mostly early-type while the weak-SFG are 
       late-type, meaning that the quenching of star formation occurs earlier than the 
       morphological transformation.       
\end{itemize}

\begin{itemize}
 \item The fraction of the weak-SFG to SF galaxies shows a different environmental dependence 
       for different stellar masses. For low-mass galaxies, there are no strong environmental effects. 
       This indicates that SF quenching occurs rapidly in low-mass galaxies, and thus they 
       have already evolved into intermediate-MXG. However, our shallow detection limit 
       at $S11$ does not allow confirming the intermediate-MXG at low mass ($<$ 10$^{10}$ M$_{\odot}$),
       except that intermediate-MXG are relatively low-mass systems among 
       massive ($>$ 10$^{10}$ M$_{\odot}$) galaxies, and likely to be located in the outer parts
       of the cluster.
\end{itemize}

\begin{acknowledgements}

This work is based on observations with $AKARI$, a JAXA project with the participation of ESA. 
This work was supported by the Korea Science and Engineering Foundation (KOSEF) grant 
No. 2009-0063616, funded by the Korea government (MEST).
JK was supported by `KASI$-$Yonsei Joint Research for the Frontiers of Astronomy and 
Space Science' program (2011) funded by Korea Astronomy and Space Science Institute.
We thank L. Piovan for providing his SED model, and Stephane Charlot and Gustavo Bruzual
for kindly sending us the new version of their CB07 model.
HSH acknowledges the support of the Centre National d\textquoteright Etudes Spatiales (CNES). 
This work is partly based on observations obtained with the MMT, a joint facility operated by 
the Smithsonian Astrophysical Observatory and the University of Arizona, and with the telescopes 
at the Kitt Peak National Observatory.
This research has made use of the NASA/IPAC Extragalactic Database (NED) which is operated by 
the Jet Propulsion Laboratory, California Institute of Technology, under contract with the 
National Aeronautics and Space Administration.

\end{acknowledgements}

\end{document}